\documentclass[prd,amsfonts,onecolumn,nofootinbib
]{revtex4}

\usepackage{graphicx}

\newcommand{\be}{\begin{equation}} 
\newcommand{\ee}{\end{equation}}
\newcommand{\bea}{\begin{eqnarray}}
\newcommand{\eea}{\end{eqnarray}}

\newcommand{\g}{ {\rm g} }

\newcommand{\gapp}{\mathrel{\raise.3ex\hbox{$>$}\mkern-14mu
              \lower0.6ex\hbox{$\sim$}}}
\newcommand{\lapp}{\mathrel{\raise.3ex\hbox{$<$}\mkern-14mu
              \lower0.6ex\hbox{$\sim$}}}

\newcommand\lsim{\lesssim}
\newcommand\gsim{\gtrsim}

\newcommand\vev[1]{{\langle {#1} \rangle}}
\renewcommand\({\left(}
\renewcommand\){\right)}
\renewcommand\[{\left[}
\renewcommand\]{\right]}

\newcommand\del{{\mbox {\boldmath $\nabla$}}}

\newcommand\eq[1]{Eq.~(\ref{#1})}
\newcommand\eqs[2]{Eqs.~(\ref{#1}) and (\ref{#2})}
\newcommand\eqss[3]{Eqs.~(\ref{#1}), (\ref{#2}), and (\ref{#3})}
\newcommand\eqsss[4]{Eqs.~(\ref{#1}), (\ref{#2}), (\ref{#3})
and (\ref{#4})}

\newcommand\eqreff[1]{(\ref{#1})}
\newcommand\eqsref[2]{(\ref{#1}) and (\ref{#2})}

\newcommand\pa{\partial}

\newcommand\mpl{M_{\rm P}}

\newcommand{\dlabel}[1]{\label{#1}}

\def\calp{{\cal P}}
\def\calr{{\cal R}}

\def\calpz{{\calp_\zeta}}
\def\calpr{{\calp_\calr}}

\newcommand\bfk{{\mathbf k}}

\newcommand\bfp{{\mathbf p}}

\newcommand\bfx{{\mathbf x}}

\newcommand\GeV{\,\mbox{GeV}}
\newcommand\MeV{\,\mbox{MeV}}

\newcommand\sub[1]{_{ \rm{#1} } }

\newcommand\mone{^{-1}}
\newcommand\mtwo{^{-2}}
\newcommand\mthree{^{-3}}

\newcommand\mfive{^{-5}}
\newcommand\mhalf{^{-1/2}}
\newcommand\half{^{1/2}}

\newcommand\mquarter{^{-1/4}}
\newcommand\threehalf{^{3/2}}

\newcommand{\fnl}{f\sub{NL}}

\newcommand{\Ai}{\mbox{Ai}}
\newcommand{\Bi}{\mbox{Bi}}

\newcommand{\dpnad}{\delta p\sub{nad}}

\newcommand{\svev}{\sub{nl}}

\begin{document}

\title{Contribution of the hybrid inflation waterfall to the primordial
curvature perturbation\footnote
{A preliminary version of this paper appeared as arXiv:1005.2461.}}

\author{David H. Lyth}
\affiliation{Department of Physics, Lancaster University, 
Lancaster LA1 4YB, UK}

\begin{abstract}
A contribution  $\zeta_\chi$ to the curvature perturbation will be  generated
during the waterfall that ends  hybrid inflation, that   may be
 significant
on small scales. In  particular, it may lead to excessive black hole formation.
We here consider  standard  
hybrid inflation, where  the tachyonic mass of the waterfall
field is much bigger than the Hubble parameter. We calculate  
 $\zeta_\chi$ in the simplest case, and see why earlier calculations
of  $\zeta_\chi$    are incorrect.  
\end{abstract}

\maketitle



\section{Introduction}

The primordial curvature perturbation $\zeta$ is one of the most important features of
the early universe. On each scale  it is relevant until the
era  of horizon entry at the epoch $k=aH$,  
and  provides the principle (perhaps the only)
  initial condition for the subsequent  evolution of all other 
perturbations.\footnote
{We use the standard cosmology notation.   The comoving wavenumber $k$ of a
Fourier component defines a comoving scale $1/k$ so that smaller scales
have bigger $k$, but $k$ itself is loosely referred to as the scale.
The physical wavenumber is $k/a(t)$ and   the 
Hubble parameter is $H\equiv \dot a/a$, while  $a_0$ and $H_0$ are evaluated at the
present epoch.
 During inflation scales `leave the horizon'
when  $aH$ exceeds $k$ and after inflation they  they enter it
when $aH$ falls below
$k$.} 
At the beginning of the known history of the universe, when the temperature is 
around $1\MeV$,  observation has established the   existence of $\zeta$
 at  wavenumbers  between
 $k \sim a_0 H_0$ (corresponding to the size of the observable universe) 
and   $k \sim e^{15} a_0 H_0$.
On these `cosmological scales' its  spectrum $\calp_\zeta(k)$
is almost scale-invariant with value $\calpz = (5\times 10^{-5})^2$.

The primordial curvature perturbation presumably exists also on much shorter scales
and at much earlier times. Although  not directly observable, this  can have 
a significant effect on the evolution of the early universe.
The most dramatic possibility is black hole formation; assuming that $\zeta$
is nearly gaussian, cosmological constraints on the abundance of primordial
black holes  requires $\calpz\lsim 10\mtwo$ at horizon entry, on practically 
all scales \cite{bhbound}.
Another important effect may occur if there is an era of matter domination in the 
early universe.  A significant  primordial
curvature perturbation  may cause the formation of
 gravitationally bound objects, with possibly dramatic  effects \cite{efg}.

The initial condition for the observable universe is presumably set by an early
era of inflation, which at the classical level makes the universe absolutely 
homogeneous. The curvature perturbation must then originate as a quantum fluctuation.

To explain the nearly
flat spectrum, the observed curvature perturbation is generally assumed to
originate from the vacuum
fluctuation of  one or more light scalar  fields during almost exponential 
inflation.\footnote
{A scalar field $\sigma$ with canonically kinetic term
is said to be light during inflation if $|m_\sigma^2| \ll H^2$, where
$m_\sigma^2\equiv \pa^2 V/\pa \sigma^2$ and $V$ is the scalar field potential,
and heavy if $|m_\sigma^2| \gg  H^2$.
A vector field perturbation might also contribute to $\zeta$ \cite{ouranis}.}
The curvature perturbation generated in this case
can exist down to the scale $k=e^{N_0} a_0 H_0$, the horizon scale at the end of inflation.
 With a   typical cosmology and inflation scale, $N_0\simeq 60$.
The 
 spectrum of the curvature perturbation generated by light field perturbations 
might increase strongly on small scales, allowing  significant
black hole formation \cite{ourbh}.

Heavy scalar fields can acquire a classical perturbation 
 through what is called preheating.
Standard preheating \cite{standardph} invokes an interaction $g^2\phi^2\chi^2$, where 
$\chi$ is the heavy field and $\phi$ is an oscillating field that is initially
homogeneous. It begins with an era, during which $\phi$
remains homogeneous so that  the field equation for $\chi_\bfk$ is
\be
\ddot\chi_\bfk =  -\( m^2(t) + k^2 \) \chi_\bfk
, \dlabel{ddotchi0} \ee
with $m^2(t) = m^2 + g^2\phi^2(t)$ and $m^2$ the mass-squared of $\chi$.
 If $m^2(t)$ is dominated by the second
term, the solution can grow exponentially, 
converting the vacuum fluctuation of $\chi_\bfk$
 to a classical perturbation. 
The contribution to the curvature
perturbation generated during this  era is calculated in \cite{ourfirstph}
assuming the inflation potential  $V\propto \phi^2$, and in 
\cite{kt} for the smooth/mutated hybrid inflation \cite{smoothhybrid}
potential.

In this paper we consider instead tachyonic  preheating, that ends 
  hybrid inflation.
Here the interaction is the same as before, but the waterfall field $\chi$
has a negative (tachyonic) mass-squared. If $\phi$ falls below a  critical value
 during inflation, $m^2(\phi)$  becomes negative and $\chi_\bfk$
can grow exponentially,  again converting its vacuum fluctuation to a classical
perturbation. To facilitate the calculation we make some fairly restrictive assumptions
about the waterfall, which will be relaxed in future papers.

The layout of the paper is as follows. In Section
 \ref{sprim} we recall the definition and properties of  $\zeta$.
In  Section \ref{shybrid} we recall the basics of hybrid inflation. 
In Section \ref{sgen}  we lay down our assumptions and describe the evolution of
waterfall field. In Section \ref{sgut}  
we define the region of parameter space in which our assumptions
are consistent.
In Section  \ref{spp}  we calculate the pressure and energy density
of the waterfall field. In  Section \ref{sduf} we justify our assumptions.
In Section \ref{vb} we calculate  $\zeta_\chi$, the contribution of the
waterfall field perturbation   to the curvature perturbation, by integrating
the perturbation theory expression for $\dot \zeta$. In 
 Section \ref{sdn} we review the  $\delta N$ approach and show that it reproduces
the previously-calculated result for $\zeta_\chi$.
In Section \ref{searlier}
we see that previous  calculations of $\zeta_\chi$  are incorrect.
In Section \ref{sconc} we conclude, pointing to the need for a better understanding
of the ultra-violet cutoff in the context of cosmological scalar field calculations.

\section{Primordial curvature perturbation $\zeta$}
\dlabel{sprim}

\subsection{Cosmological perturbations}

Let us recall some basic concepts, described for instance in \cite{book}.
Given a  function $f(\bfx,t)$ in our Universe one can  define
the perturbation $\delta f$:
\be
f(\bfx,t) = f(t) + \delta f(\bfx,t)
, \ee
where $f(t)$ refers to some  Robertson-Walker (unperturbed or background)
universe whose   line element is
\be
ds^2 = -dt^2 + a^2(t)\delta_{ij}dx^i dx^j = a^2(\eta) \( - d\eta^2 +\delta_{ij}dx^i dx^j \)
. \ee
In the background universe the energy density  $\rho$ and pressure $p$ are 
related by the continuity equation 
\be
\dot\rho = - 3H(\rho  + p)
,\dlabel{econ} \ee
where $H\equiv \dot a/a$. Also, 
Einstein gravity is assumed corresponding to the Friedmann equation
$\rho = 3\mpl^2 H^2$ with  $\mpl=(8\pi G)\mhalf =2\times 10^{18}\GeV$.

The coordinates $(\bfx,t)$ label points in the background universe, and also points
in our Universe. The latter labeling (gauge) defines slices (fixed $t$) and threads 
(fixed $\bfx$) in our Universe. Let us denote $\delta f(\bfx,t)$ in some chosen gauge
by $g(\bfx,t)$. 
We will need the first-order gauge transformation for the case that $f$ 
is specified by a single number. Going to a new coordinate system with time coordinate
$\tilde t(t,\bfx)$, it is
\bea
 \tilde g - g &=& - \dot f \( \tilde t - t \) \dlabel {gtran} \\
& =& \dot f \delta t,  \dlabel{gtran2} \eea
where $\delta t$ is the time shift going from a slice of uniform $t$, to one of uniform
$\tilde t$ with the same numerical value.

We will also need to describe the statistical  properties of $g$ at fixed $t$,
which are simplest in Fourier space;
\be
g_\bfk(t) = \int d^3x e^{-i\bfk\cdot\bfx} g(\bfx,t)
. \ee
When considering the statistical properties 
 one  ignores the spatial average (zero mode), which 
 can be absorbed into the background unless $f$ corresponds to anisotropy.

Considering an ensemble of universes, we assume that the 
the statistical  properties 
are invariant under translations
(statistical homogeneity) and rotations (statistical isotropy)
By virtue of the former, 
ensemble averages can be replaced by spatial averages within the particular 
realization
of the ensemble that corresponds to our Universe (ergodic theorem) \cite{book}.

Statistical homogeneity  implies $\vev {g_\bfk}=0$. The two-point correlator defines
the spectrum:
\be
\vev{g_\bfk g_\bfp} = (2\pi)^3 \delta^2(\bfk+\bfp) (2\pi^2/k^3) \calp_g(k)
. \dlabel{gdef} \ee
 An equivalent quantity, also called the spectrum, is $P_g\equiv (2\pi^2/k^3)
\calp_g$. 

A  gaussian perturbation is defined as one with no correlation between the
$\chi_\bfk$, except that implied by the reality condition $\chi_{-\bfk} =
\chi^*_\bfk$. Its only correlators are the two-point correlator, and
the disconnected $2^n$ point correlators starting with
\be
\vev{g_{\bfk_1} \g_{\bfk_2} g_{\bfk_3} g_{\bfk_4} }
=\vev{g_{\bfk_1}g_{\bfk_2}} \vev{g_{\bfk_3}g_{\bfk_4}}
+ \mbox{permutations} 
. \dlabel{fourpoint} \ee

Non-gaussianity of $g$
 is specified by its
bispectrum $B_g$, trispectrum $T_g$
etc., defined by
\bea
\vev{g_{\bfk_1}g_{\bfk_2}g_{\bfk_3}}
&=& (2\pi)^3 \delta^3(\bfk_1+\bfk_2+\bfk_3) B_g
, \\
\vev{ g_{\bfk_1} g_{\bfk_2} g_{\bfk_3} g_{\bfk_4} }_c
&=& (2\pi)^3 \delta^3(\bfk_1+\bfk_2+\bfk_3+\bfk_4) T_g
, \eea
etc., where the subscript c denotes the connected part of the correlator that has
to be added to the disconnected part.

The mean-square perturbation $\vev{g^2}$ can be taken as the spatial average
of $g^2(\bfx)$, and is given by
\be
\vev{g^2}=  \frac1{(2\pi)^3}\int d^3k P_g(k) = \int^\infty_0 \frac{dk}k \calp_g(k)
 \dlabel{meansq} 
 \ee
After smoothing $g$ on a scale $L$ this becomes\footnote
{A function is said to be smooth on the scale $L$ if its Fourier components
are negligible in the regime $k\gsim L\mone$. 
Smoothing means that we remove Fourier components in this regime.}
\be \vev{g^2}=\int^{L\mone}_0 \frac{dk}k \calp_g(k) \dlabel{meansq5}
. \ee

For a gaussian perturbation, the probability distribution 
of $g(\bfx)$ is gaussian so that 
 $\vev {g^n}$ vanishes for odd $n$ and is equal to $(n-1)!!(\vev{g^2}
^{n/2}$  for even n.
For a non-gaussian perturbation
 \bea
\vev{g^3} &=& (2\pi)^{-6} \int d^3k _1 d^3k_2 B_g(\bfk_1,\bfk_2), \nonumber \\
\vev{g^4}\sub c &\equiv & \vev{g^4}-3 \vev{g^2}^2
= (2\pi)^{-9} \int d^3k_1 d^3k_2 d^3 k_3 T_g(\bfk_1, \bfk_2,\bfk_3)
, \eea
etc.. The skew, kurtosis etc.\ of the probability distribution
are $S_g \equiv \vev{g^3}/\vev{g^2}^{3/2}$, $K_g
 \equiv \vev{g^4}\sub c/\vev{g^2}^2$ etc.. If they are all $\ll 1$
 $g$ can be regarded as  almost gaussian. 

We will also be interested in the spectrum of $g^2$. 
Using \eq{fourpoint}  with the convolution theorem 
\be
(g^2)_\bfk = \frac1{(2\pi)^3} \int d^3k' g_{\bfk'} g_{\bfk'-\bfk}
, \dlabel{conv} \ee
one  finds \cite{myaxion}
\be
P_{g^2}(k) 
=\frac{2}{(2\pi)^3}  \int d^3 k' P_g(k') P_g(|\bfk-\bfk'|) 
. \dlabel{calpchisq}
\ee

\subsection{Primordial curvature perturbation $\zeta$}

\dlabel{sszeta}

In the gauge with 
the comoving threading and the slicing of uniform
energy density $\rho$, the spatial metric defines $\zeta$ non-perturbatively 
through
\bea
g_{ij}(\bfx,t) &\equiv&  a^2(\bfx,t)\( e^{2h(\bfx,t)} \)_{ij},\dlabel{gij} \\
\zeta &\equiv&   \ln \(  a(\bfx,t)/a(t) \) \equiv \delta \ln a
. \dlabel{gij2} \eea
In this expression Tr\,$h=0$,
  so that the last factor has unit determinant and 
$a(\bfx,t)$ is the local scale
factor such that a comoving volume element is proportional to $a^3$. 

For this definition to be useful, we need to smooth the 
 metric and the stress-energy tensor
on some fixed   super-horizon  scale $L_\zeta$ (physical scale $L_\zeta a(t)$).
The smoothing scale must be below the shortest scale of interest, and to 
 encompass the biggest possible range of
scales it should not be  much  bigger than 
the Hubble scale at the end of inflation. Then we assume, by virtue of the
smoothing, that  spatial gradients have a negligible effect on the
 evolution of the
stress-energy tensor. This is the separate universe assumption, that is useful
also for the evolution of other quantities (we will invoke it later for scalar
fields).
The separate universe assumption will be 
 valid if relevant spatial gradients are 
 negligible, which will be the case if $L_\zeta$ is sufficiently 
far outside the horizon.  One generally 
assumes that the separate universes are isotropic (which in our case means that
the energy-momentum tensor and the expansion are isotropic)
so that the separate universes  are Robertson-Walker as 
opposed to more general Bianchi universes. 
That will be the case if scalar field 
inflation
sets the initial condition for the Universe, 
as we are assuming in this paper.
The  following treatment of $\zeta$ goes through 
however even if the separate universes
are anisotropic as might be the case if 
vector fields are involved \cite{ouranis}.

By virtue of the separate universe assumption, 
the  energy continuity equation
\eqreff{econ}  applies  at each location. 
Since  we have chosen the slicing of uniform $\rho$,
 the local continuity equation reads
\bea
\dot\rho(t) &= & - 3\frac{\dot a(\bfx,t) }{a(\bfx,t)}
\[ \rho(t) + p(\bfx,t)  \] \dlabel{dotrho} \\
&=& - 3 \[ H(t) + \dot \zeta(\bfx,t) \] \[ \rho(t) + p(t) + \delta p\sub{nad}(\bfx,t) \]  
\dlabel{dotrho2}
, \eea  
where $\dpnad$  is the pressure perturbation on the slicing of uniform density
(non-adiabatic pressure perturbation). 
Subtracting the original continuity equation
\eqreff{econ} gives
\be
\dot\zeta(\bfx,t) = -H(t)  \frac{\delta p\sub{nad}(\bfx,t)  }
{ \rho(t) + p(t) + \delta p\sub{nad}(\bfx,t) }
.\ee
During any era when $p(\rho)$ is a unique
function (the same at each location), $\delta p\sub{nad}$ vanishes and
 $\zeta$ is time-independent.

To first order in $\dpnad$ we have,
\be
\dot\zeta(\bfx,t) =  -\frac{H(t)}{\rho(t) + p(t)} \delta p\sub{nad}(\bfx,t)
\dlabel{zetadot}, \ee
with
\be
 \delta p\sub{nad}(\bfx,t) = \delta p(\bfx,t) - \frac{\dot p(t)}{\dot\rho(t)} \delta\rho(\bfx,t)
, \dlabel{deltapnad} \ee
which is valid in any gauge. Also, using \eqs{gtran2}{gij2},
\be
\zeta(\bfx,t)  = H\delta t = -H\delta\rho/\dot\rho = \frac13 \frac{\delta\rho}{\rho + p}
, \dlabel{flatdrho}  \ee
where $\delta\rho$ is the perturbation on the slicing of uniform $a(\bfx,t)$
(flat slicing).

A  second-order calculation of  $\zeta$ is   needed only to treat very small
non-gaussianity corresponding to reduced bispectrum $|\fnl|\lsim 1$.
 On cosmological scales, such non-gaussianity 
 will eventually be measurable (and is expected  if $\zeta$
comes from a curvaton-type  mechanism \cite{curvaton,curvaton2,book}).
  But there is no hope of detecting such   non-gaussianity on  much smaller scales.

Before continuing we  emphasize the following point.
A realistic Robertson-Walker universe will have  inhomogeneities,
 and one has to take their spatial average in order to 
 determine the evolution of the spatially-homogeneous 
scale factor. This is true in particular for each of the separate universes.
Various kinds of  small-scale inhomogeneity 
 could contribute to the spatially-averaged energy
density and pressure. The  generation of the classical perturbation
of light fields from the vacuum  continues 
after the smoothing scale leaves the horizon
until at least the end of inflation, on successively smaller scales. 
The effect of this perturbation on the energy density and pressure during inflation is 
removed
by the smoothing (because it is linear to high accuracy in the field perturbation)
but the light field perturbations existing at the end of inflation might have an
effect on the subsequent evolution, for instance on massless preheating 
\cite{klv}.\footnote
{In that case one may have to worry about which realization of the
ensemble of fluctuations is  the observable universe \cite{klv}.}
 There  could  be
 cosmic strings or other localized energy density configurations, whose spacing
must be much less than the smoothing scale for the  separate universe
assumption  to be valid.
Finally, through preheating,  
both heavy and light fields can acquire  classical perturbations
from the vacuum fluctuation on sub-horizon scales.
It is this effect that will be important for us, 
as it was for the ordinary preheating calculations \cite{ourfirstph,kt}.
To calculate $\zeta_\chi$, one must first calculate $\rho$ and $p$
including the sub-horizon scale field perturbations, {\em and then}
smooth $\rho$ and $p$ on  a super-horizon scale.

\section{Hybrid inflation}
\dlabel{shybrid}

In this paper we consider scalar fields with canonical kinetic terms. Their contributions
to the energy density and pressure are
\bea
\rho &=&  V  + \frac12\sum \dot\phi_i^2 + \frac12 \sum |\del \phi_i|^2 \\
p &=& - V + \frac12\sum \dot\phi_i^2 + \frac16 \sum |\del \phi_i|^2 
, \eea
where $V$ is the scalar field potential. Ignoring the metric perturbation
(back-reaction) their field equations are
\be
  \ddot \phi_i + 3H\dot\phi_i + \frac{\pa V}{\pa \phi_i} -\nabla^2 \phi_i  = 0
. \ee

\subsection{Slow-roll inflation}

We consider    single-field  slow-roll inflation, in which
the  inflaton field $\phi$ has  negligible interaction with any other field,
and is the only one with time-dependence.
Consider first the unperturbed (spatially homogeneous) universe.
The energy density and pressure are
\bea
 \rho_\phi &=& V(\phi) + \frac12\dot\phi^2 \dlabel{rhoofphi}\\
 p_\phi &=& -V(\phi)  + \frac12 \dot\phi^2,  \dlabel{pofphi},
\eea
and the field equation is 
\be  \ddot \phi + 3H\dot\phi + V'(\phi) = 0,
\dlabel{phiddot}.  \ee   
 Slow-roll inflation corresponds to almost exponential expansion:
\be
|\dot H| \ll H^2 \qquad  \Leftrightarrow \qquad \dot\phi^2 \ll \mpl^2H^2 
\qquad \mbox{(slow roll)}
, \dlabel{sr} \ee
which leads to the approximations
\be
3\mpl^2H^2 = V(\phi),\qquad 
 3H\dot\phi =  - V'(\phi)\qquad
.\dlabel{hofv}
\ee
(An over-dot means differentiation with respect to 
$t$ while a prime means differentiation with respect to the displayed 
argument.) The first derivative of \eq{sr} is also supposed to be valid, leading to
\be
|\ddot\phi|\ll H|\dot\phi|,\qquad |V''| \ll H^2
. \dlabel{vpp} \ee
For some purposes one  or more higher derivatives of \eq{sr}
are also supposed to be valid
leading to further relations, but we shall not need them.

Ignoring back-reaction, the  first-order perturbation
$\delta\phi$ satisfies 
\be
\ddot  {\delta \phi_\bfk} + 3H\dot{\delta \phi_\bfk} + \( k/ a \)^2 
 {\delta \phi}_\bfk
 = - V''(\phi(t))   \delta \phi_\bfk 
. \dlabel{ddotdeltaphi} \ee
Back-reaction  vanishes in the slow-roll limit $\dot\phi(t)\to 0$,  
 in any gauge whose slicing is non-singular in that limit \cite{book}. 
 The slicing of the widely used longitudinal 
gauge is non-singular, and so is the flat slicing \cite{book}.  
Assuming Einstein gravity,
back-reaction is a small effect in these gauges. The slicing  with uniform energy density
is singular in the limit, but $\delta\phi$ then vanishes.

As each scale leaves the horizon during slow-roll inflation, the vacuum
fluctuation generates  a nearly gaussian classical perturbation $\delta\phi_\bfk$
with spectrum $(H/2\pi)^2$.
We denote the contribution of this perturbation  to the curvature perturbation by
$\zeta_\phi$ (to  first order in the field perturbations, it is the only contribution
during single-field inflation).
After smoothing $\phi$ on a super-horizon scale,
 \eq{phiddot} is valid at each location, with the overdot meaning
$d/dt\sub{pr}$ and 
$H \equiv [ d a(\bfx,t\sub{pr}) /dt\sub{pr} ]/a(\bfx,t\sub{pr})$, 
where $t\sub{pr}$ is the proper 
time.\footnote
{To first order in $\delta\phi$, the effect of back-reaction on 
 \eq{ddotdeltaphi} takes care of the difference
between $H(t)$ and $H(\bfx,t)$ and between $t\sub{pr}$ and coordinate time
(see for instance Eq.~(5) of \cite{gwbm}).}

 By virtue of the slow-roll approximation 
$\dot\phi$ is determined by $\phi$, which means $\zeta_\phi$ is conserved. It also
means that the slicing of uniform $\rho$ is the same as the slicing of uniform
$\phi$, so that to first order in $\delta\phi$
\be
\zeta_\phi  =- H\delta\phi/\dot\phi
. \dlabel{zetaofphi} \ee
This gives 
\be
\calp_{ \zeta_\phi }(k) =  \( H^2/2\pi\dot\phi \)^2
,\dlabel{calpzetaphi} \ee
where the right hand side is evaluated at horizon exit.

\subsection{Hybrid inflation}

\dlabel{sshybrid} 

Hybrid inflation 
was proposed and named  in \cite{andreihybrid} as a way of getting the low 
inflation energy scale that might  be required by the axion
isocurvature perturbation  \cite{myfirstaxion,myaxion,axionstring}.
(See \cite{earlyhybrid} for earlier realizations of hybrid inflation.)
It was subsequently  found \cite{ourhybrid,gutinf,supernatural,runningmass} 
to be a powerful tool for model building especially in the context of
supersymmetry.\footnote
{More recently, there is interest in hybrid inflation with a non-canonical
kinetic term, in particular DBI inflation \cite{dbi}. The only essential
feature, needed to make sense of the hybrid inflation paradigm,
is that the potential dominates the energy density.}
Until  hybrid inflation nears its end
a `waterfall field' $\chi$, which has nonzero vev, 
is fixed at the origin by its interaction with $\phi$, up to a vacuum fluctuation
which is ignored. This displacement of $\chi$
from its vev gives a constant contribution to $V$  which dominates the total,
leading to single-field slow-roll inflation. After 
 $\phi$ falls below some critical value $\phi\sub c$, the waterfall field
develops a nonzero value $\chi(\bfx,t)$ which eventually ends inflation.
 That process is called the waterfall; in other words, {\em the waterfall
begins when $\phi$ falls through its critical value, and ends when inflation
ends}.

We will  adopt the  potential
\bea
V(\phi,\chi) &=& V_0 + V(\phi)  
+\frac12 m^2(\phi)  \chi^2 + \frac14\lambda \chi^4
 \dlabel{fullpot}  \\
m^2(\phi) &\equiv & g^2\phi^2  -m^2 \equiv g^2 \(\phi^2-\phi\sub c^2 \)
, \eea
with $0< \lambda\ll 1$  and $0< g\ll 1$. The potential is invoked for $\phi
\lsim \phi\sub{obs}$, the value when the observable universe leaves the horizon.
The effective mass-squared  $m^2(\phi)$ goes negative
when $\phi$ falls below $\phi\sub c\equiv m/g$, the waterfall then commencing.\footnote
{Taking account of the inhomogeneity in $\phi$, 
the waterfall begins at each location 
when  $\phi(\bfx,t)=
\phi\sub c=m/g$. Since $m\gg H$. But  the  perturbation $\delta\phi\sim H$
is small compared with $\phi\sub c$, which means that this is a small effect.}

The vev of $\phi$ vanishes
and we take $V(\phi)$ to vanish at the vev. The inflationary potential 
is supposed to be dominated by $V_0$ which means $V(\phi)\ll V_0$.
The requirements that $V$ and $\pa V/\pa \chi$ vanish in the vacuum
give the vev $\chi_0$ and the inflation scale $V_0\simeq 3\mpl^2 H^2$:
\be
\chi_0^2=\frac{m^2}\lambda,\qquad V_0=\frac{m^4}{4\lambda }\simeq 3\mpl^2 H^2 
 \dlabel{chiandv}. \ee 
The following relations  are  useful:
\be
\frac{12H^2}{m^2}=\frac{\chi_0^2}{\mpl^2},\qquad \frac{12H^2}{\mpl^2} = \lambda
\frac{\chi_0^4}{\mpl^4}
.\dlabel{chiandv2}  \ee

To justify the omission of higher powers of $\chi$  we will assume $\chi_0\ll\mpl$.
This implies $m/H\gg 1$, which is in any case 
the standard assumption for hybrid inflation.\footnote
{At least for the usually-considered potentials, this ensures that the waterfall
takes at most a few Hubble times. In the opposite case the 
 the waterfall can end with an era of two-field slow-roll inflation, involving 
$\chi$ as well as $\phi$.}
Also, to justify the omission  of high powers of $\phi$ in $V(\phi)$, 
we will assume  $\phi\sub{obs} \ll \mpl$.
  That requires
\be
 \phi\sub c\equiv m/g \ll \mpl\qquad \mbox{(small $\phi$)}
. \dlabel{smallfield} \ee
It also requires \cite{lbound} 
a tensor perturbation well below the present observational limit, implying
\be
H/\mpl \lsim 10\mfive\qquad \mbox{(tensor bound)} 
. \dlabel{tensor} \ee
Successful BBN requires an inflation scale $\sqrt{\mpl H}> \MeV$ corresponding
to 
\be
H/\mpl >10^{-42}\qquad \mbox{(BBN)} 
, \dlabel{hlower} \ee
but viable models of the early universe generally require a far higher value.

The  potential \eqreff{fullpot} was proposed 
 in \cite{andreihybrid}, with $V(\phi)=m_\phi^2 \phi^2/2$. 
If one demands $\zeta\simeq\zeta_\phi$, it
 gives spectral index $n>1$ in contradiction with observation.
 Many forms of $V(\phi)$ have been proposed that are consistent with 
$\zeta\simeq\zeta_\phi$ \cite{al,book}, and we will not assume any particular form.

Minor variants of \eq{fullpot} would make little difference to our
analysis. The  interaction $g^2\phi^2\chi^2$ might be replaced by
$\phi^2 \chi^{2+n}/\Lambda^n$ where $\Lambda$ is a uv cutoff,
or  the  term $\lambda\chi^4$ might be replaced by $\chi^{4+n}/\Lambda^n$.
 For our purpose, these  variants are  equivalent to allowing
(respectively) $g$ and $\lambda$ to be  many orders of magnitude below
unity.  
Also, $\phi$ might have two or  more 
components that vary during inflation. Most of our
 treatment of the waterfall 
 will  apply to that case,  if  $\phi$ is the field pointing 
 along the inflationary trajectory when the waterfall commences. 

More drastic modifications are also possible,
 including inverted hybrid inflation \cite{inverted}
where $\phi$ is increasing during inflation, and mutated/smooth hybrid
inflation \cite{smoothhybrid}
where the waterfall field varies during inflation. Also, the waterfall potential
might have a local minimum at the origin so that the waterfall proceeds by
bubble formation \cite{firstorder,ourhybrid}. Our analysis does not apply to those cases.

In this account of hybrid inflation we have taken $\phi$ and $\chi$ to be real fields.
 More generally
they  may correspond to directions in 
a field space that provides a representation
 of some non-Abelian symmetry group (the GUT symmetry, for the waterfall field
of GUT inflation).  That introduces some
numerical factors without changing the structure of the equations.
 For the waterfall field it 
also avoids the formation of  domain walls at locations where $\chi=0$, 
 which would be fatal to the cosmology. There  may 
instead be  cosmic strings, which are harmless if the inflation scale is
not too high. For clarity we pretend that $\phi$ and $\chi$ are real fields.

\section{Waterfall field $\chi$}

\dlabel{sgen}

\subsection{Linear era}

\dlabel{sgena}

Most  papers on the waterfall make two  basic assumptions.
First,  back-reaction
is ignored so that 
\bea
\ddot   \phi + 3H\dot \phi  -\nabla^2  \phi
& =& -   V'(\phi) - g^2\chi^2 \phi \dlabel{fullphi} \\
\ddot \chi + 3 H\dot\chi -\nabla^2\chi
&=&  - m^2(\phi) \chi - \lambda \chi^3 \dlabel{fullchi} .
\eea
Second, the waterfall is assumed to begin with an era when the last term of each 
equation is negligible; during that era, the evolution of $\chi$ is linear
and we call it the  the linear era. Since $\phi$ is smooth on the horizon scale,
its negligible evolution at each location is that of an unperturbed universe
(ie.\ its spatial gradient is negligible).\footnote
{If the waterfall takes several Hubble times, the scales on which  $\phi$ is inhomogeneous
will start to come inside the horizon but that  effect  will not be very significant.}
 Choosing a gauge whose slicing corresponds to uniform $\phi$, 
\be
\ddot \chi_\bfk + 3H \dot\chi_\bfk + \[ (k/a)^2 + m^2(\phi(t)) \] \chi_\bfk
= 0
. \dlabel{chiddot} \ee
Even if  slow-roll fails during the waterfall, the slow-roll initial condition ensures that
 $\dot\phi$ continues to be determined by  $\phi$,  which means that
the slicing is also one of uniform $\rho_\phi$ and $p_\phi$. 

The  energy density and pressure of  $\chi$ are
\bea
\rho_\chi &=&  m^2(\phi)\chi^2 + \frac12\dot\chi^2 + \frac12 |\del \chi|^2
\dlabel{rchi} \\
p_\chi &=& - m^2(\phi)\chi^2 + \frac12\dot\chi^2 + \frac16 |\del \chi|^2 
. \dlabel{pchi} \eea

\subsection{Regime of parameter space for our calculation}

\dlabel{ssassumptions}

In this paper we focus on the simplest regime of parameter space,
which will be identified in Section \ref{sgut}. In that regime  the following
conditions are satisfied;  (i) slow-roll inflation continues until the end of the 
linear era, (ii) that era takes much less than a Hubble time (iii)
the change in $\phi$ during that era is negligible.
The first two conditions imply   that the change in $\dot\phi$ is also negligible.
Under these conditions, the assumptions of the previous subsection will be justified
in Section \ref{sduf}.

Since the changes in $\phi$ and $\dot\phi$ are negligible, $m^2(\phi(t))$
decreases linearly with time. Setting
  $t=0$ at the beginning of the waterfall and using  the 
dimensionless time $\tau\equiv \mu t$ we write 
\be 
m^2(\phi(t)) = -\mu^3 t,\qquad  \mu^3 \equiv   -2g^2\phi\dot\phi = -2gm\dot\phi
. \dlabel{ourapprox} \ee

Our assumptions mean that we can ignore the 
  changes  in $H$ and $a$ during the linear era, and we  set $a=1$.
The assumption that the linear era takes much less than a Hubble time is
\be
(H/\mu)\tau\svev \ll 1\qquad (Ht\svev\ll 1)  \dlabel{secondcon}
. \ee
The assumption $\phi\simeq \phi\sub c$ corresponds to
\be
(\mu/m)^2 \ll \tau\svev\mone\qquad (\phi\svev
\simeq \phi\sub c) \dlabel{firstcon}
. \ee

We are implicitly assuming that $\chi_\bfk$, generated from the vacuum fluctuation,
can be treated as  a classical field, and we shall see that the condition for this is
$\tau\sub{nl}\gg 1$. As a result,  \eqs{secondcon}{firstcon} require
the hierarchy $H \ll \mu \ll m$.

To see when the linear era ends, let us include the small second terms
on the right hand sides of \eqs{fullphi}{fullchi}.
Since the change in $\phi$ is negligible and the perturbation $\delta\phi$ is 
 small, the unperturbed $\phi(t)$ (the spatial average)
 satisfies
\be
\ddot\phi + 3H\dot\phi \simeq 3H\dot\phi = - V'(\phi\sub c) - g^2\phi\sub c \vev{\chi^2} 
, \dlabel{unpertphi} \ee
where $\vev{\chi^2}$ is the spatial average of $\chi^2$. 
Subtracting this from the full equation for $\phi$ we find
\be
\ddot{\delta\phi} + 3H\dot{\delta\phi} - \nabla^2\delta\phi
= -V''(\phi\sub c)\delta\phi - g^2 \chi^2 \delta\phi - g^2 \phi\sub c 
\delta\chi^2 , \dlabel{deltaphi2}
\ee
where
\be
\delta\chi^2 \equiv \chi^2 - \vev{\chi^2}
. \dlabel{deltachisq} \ee
Finally, \eq{fullchi} becomes
\be
\ddot \chi + 3 H \dot \chi - \nabla^2 \chi =
 - m^2(\phi(t)) \chi - \lambda \chi^3 - \delta (m^2 ) \chi 
\dlabel{chi2}
 \ee
where $\delta (m^2) \equiv m^2(\phi(\bfx,t)) - m^2(\phi(t))$.
The linear era ends when the first term on the right hand side 
 ceases to dominate, for one of these  equations.
For  \eq{unpertphi} this happens  at roughly the epoch when
$\vev{\chi^2}$ has the value 
\be
\chi\svev^2 \equiv  3H|\dot\phi|/gm
. \dlabel{chisvev} \ee
The   constraints \eqsref{sr}{smallfield} imply  $\chi\svev \ll \chi_0$,
 which means that the second term on the right hand side 
of \eq{chi2} is smaller than the first one  at this epoch.
We will show  in Section \ref{sduf} that the last terms  of 
\eqs{chi2}{deltaphi2} can also be ignored,  which means that
 $\chi\svev$ marks the end of the linear era.
Afterward, the last term of \eq{unpertphi} will cause $\phi(\bfx,t)$
to quickly decrease, and $m^2(\phi)$ will quickly move to its final value
ending inflation and (by definition) the waterfall.

\subsection{Solution of the waterfall  field equation}

\dlabel{iva} 

By virtue of our condition (ii) we can ignore the expansion of the universe,
 setting $H=0$ in  \eq{chiddot} to get
\be
\frac{d^2 \chi_\bfk(\tau)}{d\tau^2}
= x(\tau,k) \chi_\bfk(\tau),\qquad x\equiv   \tau-   k^2/\mu^2
=-\( m_\phi^2(\tau) + k^2 \)/\mu^2
. \dlabel{modefu}\ee
The solutions  are the Airy functions $\Ai(x)$ and $\Bi(x)$. 

In the  quantum theory, $\chi_\bfk$ becomes an operator $\hat\chi_\bfk$.
Working in the Heisenberg picture we write
\bea
\hat \chi_\bfk(\tau) &=& \chi_k(\tau) \hat a_\bfk + 
\chi_k^*(\tau) \hat a_{-\bfk} \\
\[ \hat a_\bfk, \hat a_\bfp \] &=& (2\pi)^3 \delta^3(\bfk-\bfp)
. \eea  
The mode function $\chi_k$ satisfies \eq{modefu}
and its Wronskian is  normalized to $-1$.  We choose
\be
\chi_k =\sqrt{\pi/2\mu} \[ \Bi(x) + i \Ai(x) \]
. \dlabel{chik1} \ee
Then, in the early-time regime $x\ll -1$, 
$E_k^2 \equiv m_\phi^2(\tau)+k^2$ is slowly varying
($|\dot E_k|\ll E_k^2$) and the theory 
 describes particles with mass-squared $m_\phi^2(t)$. Indeed, the solution
in this regime is
\be
\chi_k(\tau) = (2\mu)\mhalf |x|\mquarter e^{-i\pi/4} 
e^{-\frac23 i |x|\threehalf}
, \dlabel{chik2} \ee
which can be written
\be
\chi_k(t) = \frac1{\sqrt{2E_k}} e^{-i\int^t dt E_k}
. \ee
 We choose the 
 state vector as the  vacuum, such that $\hat a_\bfk|>=0|>$. A 
significant occupation
number is excluded
since the resulting positive pressure would spoil inflation 
\cite{areview}.

\subsection{Classical waterfall field $\chi(\bfx,t)$}

We will assume  that $\tau$ becomes much bigger than  1 during the linear era.
Then  there exists a  regime 
  $x\equiv \tau-(k/\mu)^2 \gg 1$ in which 
\be
\chi_k(\tau) \simeq  (2\mu)\mhalf x\mquarter e^{\frac23 x^{3/2}},\qquad
\dot\chi_k \simeq \mu \sqrt x  \chi_k
,  \dlabel{latemode} \ee
the errors vanishing in the limit $x\to\infty$. In this regime
 $\hat \chi_\bfk(\tau)=  \chi_k(\tau) ( \hat a_\bfk + \hat a_{-\bfk})$ 
to high accuracy. As a result, $\hat\chi_\bfk$ is  a practically
constant operator times a c-number,
which means
 that $\chi_\bfk$ is a classical quantity
in the WKB sense.
By this, we mean  that a suitable
measurement of $\chi_\bfk$ at a given time will give a state
that corresponds to a practically definite value $\chi_\bfk$ far 
into  future  \cite{book}.
(We have nothing to say about the cosmic
Schr\"odinger's Cat problem that now presents itself.) After the measurement,
$\chi_\bfk(\tau)$ is classical and
\be
\chi_\bfk(\tau) \propto \chi_k(\tau)
\dlabel{wed2} . \ee

{}Keeping only the classical modes $\chi_\bfk$, we generate a classical field
 $\chi(\bfx,t)$. 
Its spatial average vanishes and we can treat it as a gaussian perturbation.
Its spectrum  is defined by \eq{gdef}, with the  ensemble corresponding  to the 
 different outcomes of the measurement that is  supposed
to have been made to produce $\chi_\bfk(t)$. The expectation value in
\eq{gdef} can therefore be identified with the vacuum expectation value
$\vev{\hat\chi_\bfk\hat\chi_\bfp}$, which gives 
$P_\chi(k,\tau) = \chi_k^2(\tau)$. The classical mode function $\chi_k(\tau)$ is
nonzero only in the regime $x\equiv \tau-k^2/\mu^2 \gg 1$, where it is given by
\eq{latemode}. At $k=0$ we have
\be
P_\chi(0,\tau) = \chi^2_{k=0}(\tau) = (2\mu\tau\half)\mone e^{ \frac43 \tau^{3/2} }
. \dlabel{pchi0} \ee
At fixed $\tau$, $P_\chi$ is a decreasing function of $k$. 
At $k^2\ll \mu^2\tau$,
\bea
P_\chi(k,\tau) &=&\chi^2_k(\tau) \simeq \chi_{k=0}^2(\tau)  
e^{-2\sqrt \tau  k^2/\mu^2} \dlabel{specdef2} \\
&= & \chi_{k=0}^2(\tau)   e^{-k^2/k_*^2(\tau)},\qquad  
k_*^2(\tau)\equiv  \mu^2/2\sqrt\tau \dlabel{specdef}
, \eea
which means that $P_\chi(k)$ 
 decreases exponentially in the regime $k_*^2(\tau)\ll k^2 \ll \mu^2 \tau$.
Modes with $k\gg k_*(\tau)$   can therefore be ignored.
The  dominant modes
(corresponding to the peak of $\calp_\chi$ at fixed $\tau$) have
 $k\sim k_*(\tau)$.  {}From \eq{secondcon} we have $k_*(\tau)\gg H$ which means
that the dominant modes are sub-horizon. 

{}From  \eq{meansq}
 we see that\footnote
{The argument $\tau$ in $\vev{\chi^2(\tau)}$ is inserted to remind us 
 that the expectation value
of $\chi^2$ (defined as the ensemble average or equivalently the spatial average for
a given ensemble) depends only on time and not on position.} 
\be
\vev{\chi^2(\tau)}\sim P_\chi(0,\tau) k_*^3(\tau)
\dlabel{roughchisq} . \ee
 Doing the integral we find
\be
\vev{\chi^2(\tau)} = \frac{4\pi}{(2\pi)^3}
 P_\chi(0,\tau)\int^\infty_0 dk k^2 e^{-(k^2/k_*^2(\tau))}
=(2\pi)^{-3/2}
 P_\chi(0,\tau) k_*^3(\tau) 
. \dlabel{classvev} \ee

{}From \eqs{pchi0}{classvev},  $\tau\svev \sim  \( \ln(\chi\svev/\mu) \)^{2/3} $,
with $\chi\svev$ given by \eq{chisvev}. 
To have a classical era $\tau\gg 1$ we require $\tau\svev \gg 1$.
To get an upper bound on $\tau\svev$ we use $\chi\sub{nl} \ll 
 \chi_0\ll \mpl $ with $\mu\gg H$, to 
find $\tau\svev \lsim \(\ln (\mpl/H) \)^{2/3}$. 
According to \eqs{tensor}{hlower}, this upper bound on $\tau\svev$
is of order $5$ to $20$ with the lower end of the range far more likely. 

Since the dominant modes have $k^2\sim k_*^2(\tau)\ll \mu^2\tau$, we can set
$x\simeq \tau$ to get $\dot\chi_k\simeq \mu \sqrt \tau \chi_k $. 
At a typical location we have to a good approximation
\be
\dot\chi(\bfx,\tau) = \mu \sqrt\tau \chi(\bfx,\tau)
. \dlabel{chidot}
\ee 
We also have
\be
\vev{|\del\chi|^2} \sim \int d^3k k^2 |\chi_k|^2 \sim k_*^2(\tau) \vev{\chi^2}
\ll \vev{\dot\chi^2} =\mu^2\tau \vev{\chi^2}
. \dlabel{gradchi} \ee
At a typical location, 
  $|\del\chi|$ is of order $k_* \chi$, which is  negligible
compared with the typical time-derivative $\mu\sqrt \tau \chi$.
Integrating \eq{chidot} over an extended time interval  gives
\be
\chi(\bfx,\tau) \propto e^{(2/3)\tau^{3/2} }
. \dlabel{rough} \ee
According to \eqss{pchi0}{chidot}{gradchi}, this approximation ignores the
time-dependence of $k_*$ (and less seriously, the prefactor in \eq{pchi0}).
That approximation is good over a time interval $\Delta \tau\ll \tau$,
and we will  use  it over such an 
interval ending at $\tau\simeq\tau\svev$.  

These results may not hold  near places where
 $\chi(\bfx,\tau)$ vanishes, because there is then a cancellation between
different Fourier components and the relation $\dot\chi_k \simeq \mu \sqrt\tau
\chi_k$ cannot be expected to give \eq{chidot}. 
The places where $\chi$ vanishes are those in which 
 cosmologically disastrous domain walls would form if $\chi$ were really
a single field  and where cosmic strings might form in 
a more realistic case.
At a typical location $\del\chi/\chi$ is of order  $k_*$, therefore
the typical spacing between
these places is of order $k_*\mone$. But at a distance $L$ from one of the 
places,  $\chi$ is typically of order $L|\del \chi|\sim
L k_* \vev{\chi^2}\half$, which is well below its typical value
only within a small region $L\ll k_*\mone$. Outside this region, there is no
 strong cancellation between  Fourier components, and we may expect the 
approximations to be valid.

\section{Constraints on the parameter space}

\dlabel{sgut}

In this section we 
 identify the part of parameter space in which our assumptions are consistent.
There are  four independent parameters, which we will
take  to be the  dimensionless quantities\footnote
{Strictly speaking, we
 cannot  consider $\zeta_{\phi}(k\sub{end})$, because
$\zeta$ is usefully defined only after smoothing on a super-horizon scale.
But  the  analogue of $\zeta$ defined on the slicing  orthogonal to comoving worldlines
(usually denoted by $\calr$) remains useful with a smaller smoothing scale
and \eq{fdef} as well as the black hole bound are  valid  with $\calpz$ replaced by
$\calpr$ \cite{bhend}.}
\be
g\ll 1,\qquad H\sub P\equiv H/\mpl \ll 1,\qquad H_m\equiv H/m\ll 1, \dlabel{beforefdef}
\ee and
\be   f \equiv 
 \(5\times 10^{-5}\)\mone H^2/2\pi\dot\phi  = \(5\times 10^{-5}\)\mone
\calp_{\zeta_\phi}\half(k\sub{end} )
. \dlabel{fdef} \ee
 
Inflation models are usually constructed so that $ \calp_{\zeta_\phi}$
accounts for the observed  $\calpz$ on cosmological scales. Then, if
$\calp_{\zeta_\phi}$  is nearly scale-independent we will have
$f\sim 1$. More generally there is an upper bound
\be  f\lsim 2\times 10^3 \dlabel{bhole}
 \ee
corresponding to  the bound $\calp_{\zeta_\phi} \lsim 10\mtwo$ that is required to avoid
excessive black hole production from  $\zeta_\phi$.\footnote
{Since $\phi$ and $\chi$ have independent vacuum fluctuations, $\calpz$
is the sum of $\calp_{\zeta_\phi}$ and
$\calp_{\zeta_\chi}$,  and the black hole bound applies to each of them.}

In terms of these four parameters, useful relations are
\bea
\tau\svev \sim \[ \ln (H_m g\mone f^{-1/5}) \]^{2/3}, \\
\mu^3 \sim 10^4 gH^3 H_m\mone f\mone
\dlabel{useful} , \eea
and  the various  constraints become
\bea
10^{-42} &<& H\sub P \lsim 10^{-5}\qquad \mbox{(BBN \& tensor)}
\dlabel{bbn} \\
H\sub P &\ll & 10\mthree f \qquad \mbox{(slow roll)} \dlabel{sr2} \\
g H\sub P &\ll& H_m  \qquad \mbox{(small $\phi$)} \dlabel{smallchi2} \\
H\sub P &\ll& H_m^2\qquad  (\lambda \ll 1) \\
(\tau\svev/10)^3 H_m f &\ll&  g \qquad (Ht\svev \ll 1) \dlabel{third} \\
10^3 \tau\svev^{3/2} g H_m^2 &\ll & f  \qquad (\phi\svev \simeq\phi\sub c)
\dlabel{fifth}  \\
gf^{1/5}   &\ll& H_m \qquad (\tau\svev \gg 1) \dlabel{fourth}  \\
\eea

The constraint $\tau\svev\gg 1$ corresponds to the existence of a classical
regime, which fails if $g f^{1/5} \gsim H_m$.
As seen in Section \ref{iva}, there is an absolute bound
$\tau\svev \lsim 20$ with  $\tau\svev \lsim 10$ more likely.
\eqs{third}{fourth} require $f\ll (10/\tau\svev)^3$ which is stronger
than \eq{bhole}. If this bound on $f$ is saturated,   \eqs{third}{fourth}
become  $g H_m\mone\sim 1$ and then \eq{fifth} becomes
$g\ll (10\tau\svev\half)\mone$. For smaller $f$  the range
for $g H_m\mone$ becomes wider and \eq{fifth} bounds the product
$g H_m^2$.

\section{Pressure and energy density of $\chi$}

\dlabel{spp}

Taking the spatial gradient to be negligible (ie.\ excluding places where
$\chi$ is close to zero),  \eqss{rchi}{pchi}{chidot} give
\bea
\rho_\chi(\bfx,\tau) &\simeq& -\frac12 \mu^2 \tau \chi^2(\bfx,\tau) + 
\frac12\dot\chi^2(\bfx,\tau) \simeq 0
 \dlabel{rhoofx} \\
 p_\chi(\bfx,\tau) &\simeq & \frac12 \mu^2 \tau \chi^2(\bfx,\tau) + 
\frac12\dot\chi^2(\bfx,\tau) \simeq \dot\chi^2(\bfx,\tau)\simeq 
\mu^2 \tau \chi^2\equiv  |m^2(\phi)|^2\chi^2
. \dlabel{pofx} \eea

The cancellation of the terms in the first expression occurs because we have ignored the 
expansion of the universe when calculating $\chi$, so that energy conservation prevents 
$\rho_\chi$
moving away from its initially zero value. Instead of taking the expansion into account 
directly,
we  invoke the local energy continuity equation for $\rho_\chi$,
 which will be valid to a good approximation because the gradient of $\chi$ is negligible.
This gives $\dot\rho_\chi(\bfx,t) = - 3H p_\chi(\bfx,t)$. 
Using \eq{rough} to integrate $\dot\rho$
gives\footnote
{\eq{rough} will be a good approximation because the integral is 
dominated by values of $t$ close to 
$t\svev$.}
\be
\rho_\chi(\bfx,t) = - \frac{3H}{2\mu\sqrt \tau}p_\chi(\bfx,t)
 = -\frac{3Ht}{2\tau\threehalf}  p_\chi(\bfx,t) \dlabel{rhochi}
. \ee
Since  $Ht\svev\ll 1$ and $\tau\gg 1$, this gives $|\rho_\chi| \ll |p_\chi|$ 
as is required for consistency. 

To a good approximation the    spatially averaged pressure is 
\be
\vev{p_\chi(\tau)} = \mu^2 \tau  \vev{\chi^2(\tau) } = \vev{\dot\chi^2}
 \dlabel{papprox2} 
 \ee
and the pressure perturbation is
\be
\delta p_\chi(\bfx,\tau) = 
 \mu^2 \tau \delta\chi^2(\bfx,t) = \delta \dot \chi^2 
\dlabel{deltap} , \ee
where 
\be 
 \delta\chi^2(\bfx,\tau) 
\equiv  \chi^2(\bfx,\tau) - \vev{\chi^2(\tau)},\qquad
\dot\chi^2(\bfx,\tau) 
\equiv  \dot\chi^2(\bfx,\tau) - \vev{\dot\chi^2(\tau)}
. \ee
Multiplying \eqs{papprox2}{deltap} 
 by $-3H/2\mu\sqrt\tau$ we arrive at the corresponding expressions for
$\rho_\chi$.

Our goal is to evaluate the contribution of $\chi$ to the curvature perturbation $\zeta$, which 
depends linearly on $\delta p_\chi$ and $\delta \rho_\chi$, and hence on $\delta\chi^2$. 
Since $\zeta$ is  by definition  smooth on a super-horizon scale we are interested only in $\delta\chi^2$
smoothed on such a scale.  Let us determine
its statistical properties, starting with its spectrum.

Super-horizon  scales  satisfy {\em a fortiori}
 $k\ll k_*(\tau)$,  which means that we can
 set $\bfk=0$  within the integral \eqreff{calpchisq}. This gives roughly 
\be 
P_{\delta \chi^2}(\tau,k)  \sim P_\chi^2(\tau,0) k_*^3(\tau)
. \dlabel{papprox} \ee
Doing the integral and using \eq{classvev}, and going to $\calp_{\delta\chi^2}$,
we find
\be \calp_{\delta\chi^2}(\tau,k) = \frac1{\sqrt\pi} \vev{\chi^2(\tau)}^2 [k/k_*(\tau)]^3
. \dlabel{pchisq} \ee 
Taking the smoothing scale to be the horizon  scale, \eq{meansq5} gives
\be
\frac { \vev{ (\delta\chi^2)^2 }  }{ (\vev{\chi^2})^2 }
= \frac13 \frac{ \calp_{\delta\chi^2}(H) }{ (\vev{\chi^2})^2 }
 \sim (k_*(\tau)/H)\mthree  \ll 1
, \dlabel{monday61} \ee
and for a bigger smoothing scale $L$ we should replace $H$ by $L\mone$, making the
bound tighter. We see that {\em after smoothing on a super-horizon scale,
$\chi^2$ is almost homogeneous.}

Using the convolution theorem 
one finds at $k\ll k_*(\tau)$ the scale-independent 
expressions 
\be
B_{\delta\chi^2}(\tau) \sim P^3_\chi(\tau,0) k_*^3(\tau),\qquad
T_{\delta\chi^2}(\tau)   
 \sim  P_\chi^4(\tau,0) k_*^3(\tau)
, \ee
etc.\  
After smoothing $\delta\chi^2$ on the horizon  scale,
\be
\vev{ (\delta\chi^2)^3 } 
 \sim P_\chi^3 k_*^3(\tau) H^6,\qquad  
\vev{ (\delta\chi^2)^4 }\sub c
\sim P_\chi^4 k_*^3(\tau) H^6
, \ee
etc.. {}From these follow the 
 skew, kurtosis etc.\ of the probability distribution of $\chi^2$,
which are seen to be small:
\be
S_{\delta\chi^2} 
\sim (k_*(\tau)/H)^{-3/2}\ll 1,\qquad
K_{\delta\chi^2} 
\sim ( k_*(\tau)/H)\mthree \ll 1
. \ee
We conclude that {\em after smoothing on a super-horizon scale, $\delta\chi^2$
is almost gaussian.} 

In all of this, we have taken the  spatial average of  $\chi$ to vanish. That is
 is true  within an  indefinitely large region because $\chi$ is supposed to be 
generated entirely from the vacuum fluctuation.
But to  make contact with observation we  should  consider a finite box,  
 that is not many orders of magnitude bigger than the
region in which the observations are made \cite{mybox}.
Denote the average within the box  by  $\bar\chi$ we have
\be
\chi(\bfx) = \bar\chi + \chi\sub{calc}(\bfx)
 \ee
where the second term is the one we calculated. This gives
\be
\delta\chi^2 = \delta\chi^2\sub{calc} + 2 \bar\chi \chi\sub{calc}
. \ee
For our results to be valid we need 
$\vev{ ( \delta\chi^2\sub{calc}})^2$ to be much bigger than 
$\bar\chi^2 \vev{ \chi^2\sub{calc} } $,
when $\delta\chi^2\sub{calc}$ and $\chi\sub{calc}$ are smoothed on
any scale $k\mone$ much smaller than the box size $L$.
The  expected value of  $\bar\chi^2$  for a random location of the 
box is $\vev{\chi^2\sub{calc} }$ where now $\chi^2\sub{calc}$ is 
smoothed on the scale $L$. We therefore expect 
\be
\frac{ \vev{ ( \delta\chi^2\sub{calc} )^2 }  }{
\bar\chi^2 \vev{ \chi^2\sub{calc} } } \sim
\frac { \calp_{\delta\chi^2} (k) } {   \calp_\chi(L\mone) \calp_\chi(k) }
\sim (kL)^3 \gg 1
. \ee
We conclude that the spatial average $\bar\chi$ is negligible, unless
we live at a a very untypical location.

\section{Justifying our assumptions}

\dlabel{sduf}

\subsection{Neglect of back-reaction}

\dlabel{sbackr}

According to the  calculation of 
$\chi$ in Section  \ref{sgen}, the expansion of the universe
is negligible and    
the dominant modes of $\chi$  have $k \sim k_* \gg H$. Therefore, the calculation can be 
regarded as a flat spacetime calculation, formulated at each epoch within a locally
inertial frame 
corresponding to a box whose size $L$ satisfies  $H\mone \ll L\ll k\mone$,
 that is at rest with respect to the cosmic fluid. 
Since spatial gradients are negligible, each element of the 
 cosmic fluid is free-falling.

Since the calculation within each box takes place in practically flat spacetime,
back-reaction will have practically no effect on it.
It follows that our calculation is valid globally, in the gauge with free-falling 
threads, and with the time coordinate corresponding to proper time along 
each thread (a  synchronous gauge). 

\subsection{Neglected  terms  in \eqs{deltaphi2}{chi2}}

Let us restore the neglected terms in
\eqs{deltaphi2}{chi2}\footnote
{We thank Neil Barnaby for pointing out the need to consider these.}
We  write $\delta\phi=\delta\phi_1 + \delta\phi_2$ and
$\chi=\chi_1+\chi_2$, where $\delta\phi_1$ and $\chi_1$ are the solutions
of \eqs{ddotdeltaphi}{chiddot}, that we considered earlier. Then 
\eqs{deltaphi2}{chi2} give 
\bea
\ddot  {\delta \phi_2} + 3H\dot{\delta \phi_2}  -\nabla^2 {\delta \phi_2}
& =& - \[ V''(\phi(t)) + g^2 \chi_1^2(t) \] 
\delta\phi_2 - g^2 \phi\sub c \delta\chi^2,   \dlabel{deltaphi22} \\
\ddot \chi_2 + 3 H \dot \chi_2 - \nabla^2 \chi_2 &=&  2g^2 \phi\sub c \chi_1 \delta\phi_2
\dlabel{chi22}
. \eea
where $\delta\chi^2\equiv \chi_1^2-\vev{\chi_1^2}$ as before.
In writing these equations, we assumed $|\delta\phi_2| \ll \phi\sub c$
and $|\chi_2|\ll |\chi_1|$, 
and we will verify the self-consistency of these 
assumptions. 

Recall that $(\delta \chi_1^2)\dot{}= 2\mu \sqrt \tau \delta \chi_1^2$
and  $(\delta \chi_1^2)\ddot{}= 4\mu^2  \tau \delta \chi_1^2$.
Also, $|V''|\ll H^2$, $H\ll \mu$
and $g^2\chi^2\svev \sim g^2 H\mu^3/m^2\ll \mu^2 $. It follows that the 
first term on the right hand side of 
 \eq{deltaphi22} can be ignored, and that its solution is
\be
\delta\phi_2 = -\frac{g^2}{4\tau} \frac{\phi\sub c}{\mu^2} \delta\chi^2
,  \ee
Putting this into \eq{chi22} and remembering that $\chi_1\propto
\exp[ (2/3) \tau^{3/2} ]$ we get
\be
\chi_2 = 2 \frac{g^2}\tau  \phi\sub c \chi_1 \delta\phi_2/\mu^2
. \ee
Using $|\delta\chi|^2 \sim |\chi_1|^2< \chi^2\svev$ one easily verifies that
$|\delta\phi_2| \ll \phi\sub c$ and $|\chi_2|\ll |\chi_1|$.
The neglect of back-reaction in \eqs{deltaphi22}{chi22} is justified because the dominant
mode of $\delta\phi_2$ and $\chi_2$ is sub-horizon,  being the dominant mode
$k_*$ of  $\chi_1$.

Since $\chi_2$ is much less than $\chi_1$, it has a negligible effect on $\zeta_\chi$.
The effect of $\delta\phi_2$ is also negligible, because its  contribution
$\delta p_2=-V' \delta \phi_2$ to $\delta p$ 
is small compared with that of $\delta \chi$:
\be
\frac{ | \delta p_2| }{ |\delta p_\chi | } = \frac3{8\pi^2}  \frac H \mu \ll 1
. \ee

\subsection{Justifying the assumption of a linear era}

The above discussion shows that the assumption of a linear waterfall era is 
self-consistent, and that such an era will end only at the epoch
$\vev{\chi^2}=\chi^2\svev$ when the last term of \eq{unpertphi} becomes
significant. 
What about the possibility that, although self-consistent, the
assumption is actually wrong? For the assumption to hold, it is essential
that   $V'$ dominates the evolution during some initial era of the waterfall.
 In the opposite case where $V'$ is completely negligible, 
 both $\phi$ and $\chi$ will be  generated entirely by the potential
$\frac12 g^2(\phi^2-\phi\sub c^2)\chi^2$. This case is discussed in \cite{dufaux}.
Noting that the most rapid growth will
be in the direction of steepest descent of the potential, they find 
$\sqrt2 (\phi\sub c - \phi) \simeq \chi$ and
\be
V = V_0 - \frac{gm}{\sqrt 2} \chi^3
. \ee

This potential leads to a non-linear equation for $\chi(\bfx,t)$, which has
been considered in \cite{tachyonic}. To handle it, they
 give $\vev{\chi^2}$ an initial value coming from its vacuum fluctuation. Of course
that requires  a prescription for handling the uv divergence of $\vev{\chi^2}$ 
that is different from ours which is to simply drop modes that are in the quantum
regime. (We comment
on this  issue  at the end of Section \ref{sconc}.)

The authors of \cite{tachyonic} 
 find that $\chi$ grows like $\exp(k\sub{quant} t)$ with
$k\sub{quant}\sim gm$.
This growth rate may be compared with the growth rate $\chi\propto
\exp[\frac23 (\tau)^{3/2}]$ that we found by assuming the existence
of an initial linear era, where $\tau\equiv \mu t$ with $\tau\gg 1$.
The latter growth rate will be bigger than  the former if
 $gm\ll \mu \tau\half$, and \cite{dufaux} suggest that this
will  be the criterion for the linear era to exist. That criterion is satisfied
with our assumptions, because \eqss{useful}{bhole}{fourth} imply  $gm\ll  \mu$. 
We conclude that the assumption of an initial linear 
era has some justification, though there is as yet no conclusive proof. 

\section{The waterfall contribution to $\zeta$}

\dlabel{vb}

\subsection{Calculating the waterfall contribution}

The  contribution to $\zeta$ generated at a given epoch during the  
waterfall is
\be
\zeta_\chi(\bfx,\tau) \equiv \zeta(\bfx,\tau) - \zeta(\bfx,\tau_1)
, \dlabel{zetachidef} \ee
where $\tau_1\sim 1$ is the epoch when $\chi$ becomes classical.  We are interested in 
$\zeta_\chi(\bfx,\tau\svev)$.

Following the procedure of \cite{ourfirstph} used for ordinary preheating, let us calculate
$\zeta_\chi$ by integrating $\dot\zeta$. 
Since $\zeta_\phi$ is constant, $\delta p\sub{nad}$  receives no
contribution from $\delta \rho_\phi$. 
Using \eqsss{rhoofphi}{pofphi}{pofx}{rhochi} and remembering that $|\rho_\chi|
\ll |p_\chi|$, we find
\bea
\delta p\sub{nad} &=& \delta p_\chi - \frac{\dot p}{\dot \rho} \delta \rho_\chi
\dlabel{pnad1} 
  \\
&=& \frac{\dot\phi^2 }{ \dot\phi^2 +  \vev{\dot\chi^2 }} {\delta p_\chi} \dlabel{pnad2}
. \eea
Using this result and invoking \eq{rough}, this gives 
\be
\zeta_\chi(\bfx,\tau\svev) = -\frac H\mu \int^{\tau\svev}_{\tau_1} 
\frac{\delta p\sub{nad}(\bfx,\tau) }
{ \vev{\dot\chi^2(\tau)} + \dot\phi^2  } d\tau =
\frac13 
\frac{ \delta \rho_\chi(\bfx,\tau\svev)}{\vev{\dot\chi^2(\tau\svev)} + \dot\phi^2  } 
-
\frac13 
\frac{ \delta \rho_\chi(\bfx,\tau_1)}{\vev{\dot\chi^2(\tau_1)} + \dot\phi^2  } 
, \dlabel{zetachibest1} \ee
where
\be
\frac13 
\frac{ \delta \rho_\chi(\bfx,\tau)}{\vev{\dot\chi^2(\tau)} + \dot\phi^2  } 
= - \frac H{ 2\mu \sqrt{\tau} } \frac{ \delta \chi^2(\bfx,\tau) }
{ \vev{\chi^2(\tau)} }
\frac{ \vev{ \dot\chi^2(\tau) } }
{ \vev{ \dot\chi^2(\tau) } + \dot\phi^2  } 
. \ee

{}From \eqs{chisvev}{papprox2} we find 
\be
 \vev{\dot\chi^2(\tau\svev) }  /\dot\phi^2 = 6 H t\svev
. \ee
We are assuming   $Ht\svev \ll 1$, which suggests $6H t\svev\ll 1$
hence   $\vev{\dot\chi^2(\tau\svev)}\ll \dot\phi^2$.  Even if $6H t\svev$
is  somewhat bigger than 1,
 $\vev{\dot\chi^2}\ll \dot\phi^2$ until $\tau$ is close to $\tau\svev$; indeed,
from \eq{rough} we have  $\vev{\dot\chi^2}\sim  \dot\phi^2$ at
the epoch $\tau_=$ given by
\be
\tau\svev - \tau_=  \sim \ln[\vev{\dot\chi^2\svev}/\dot\phi^2]/2\tau\svev\half
\ll \tau\svev
. \dlabel{teq} \ee
It follows that the integral \eqreff{zetachibest1} is dominated by the  range
$\tau\svev -\tau\ll \tau\svev$, which means that \eq{rough} 
will be a good approximation. In either case, the integral of \eq{zetachibest1}
is dominated by its upper limit so that to a good approximation
\be
\zeta_\chi(\bfx,\tau\svev) =
\frac13 
\frac{ \delta \rho_\chi(\bfx,\tau\svev)}{\vev{\dot\chi^2(\tau\svev)} + \dot\phi^2  } 
. \dlabel{zetachibest} \ee

In  the 
regime  $6Ht\svev \gg 1$ we have the following interesting result.\footnote
{That  regime can hardly exist with our restriction $Ht\svev \ll 1$, but it turns 
out \cite{mynext} that the result still holds in a regime  $Ht\svev\gsim 1$.}
Here, \eq{pnad1} becomes
\be
\delta p\sub{nad} = \delta p_\chi - \frac{\dot p_\chi}{\dot \rho_\chi} \delta \rho_\chi
, \ee
with $|\delta p\sub{nad}| \ll |\delta p_\chi|$. 
Therefore   $p_\chi$ is (almost) uniform on the slice of uniform  $\rho$
(practically coinciding with the slice of uniform $\rho_\chi$). That in turn implies that
{\em $\chi^2(t,\bfx)$ is almost uniform on the slice of uniform $\rho$}. 

The spectrum of $\zeta_\chi(\bfx,\tau\svev)$  can be written
\be
 \calp_{\zeta_\chi}(k) = 
\frac{36}{ \sqrt{2\pi} } \frac { \tau\svev^{-21/4} \( H t\svev \)^7 } 
{ \(1+ 6 H t\svev  \)^2 }  \( \frac kH \)^3
. \dlabel{calpzchi} \ee
This is the main result of our paper. It  holds 
 only in the regime of parameter space for  which $Ht\svev\ll 1$, and 
only for  $\tau\svev \gg 1$ (so that $\chi$ is classical). 
Also, since $\zeta$ is defined only after smoothing on a super-horizon scale,
it holds only for $H\gg k$.  Requiring say $(Ht\svev)\mone$,
$\tau\svev$ and $H/k$ all bigger than  2 we see that 
 $\calp_{\zeta\chi}$ is far below the black hole bound $\calpz\sim 10\mtwo$.

Taken to apply at say  $(H/k)^3=10^2$, the black hole bound implies
 on a 
 scale $k= e^{-N}H$ leaving the horizon $N$ Hubble times before the end
of inflation $\calp_{\zeta_\chi} \lsim e^{-3N}$.
Considering the shortest cosmological scale, 
 the  observed value $\calp_\zeta(k)\sim 10^{-9}$ requires
 $N<7$  which is unlikely  with any reasonable 
post-inflationary cosmology even  with a 
 low inflation scale. We conclude that the black hole constraint on the scale
$k=H$ almost certainly makes $\calp_{\zeta_\chi}$  negligible on cosmological scales.

\section{The $\delta N$ approach}

\dlabel{sdn}

The calculation of $\zeta_\chi$ in the previous section
uses  cosmological perturbation theory.
For the contributions of light field perturbations to $\zeta$,
the $\delta N$ approach  provides a powerful alternative making no reference  to cosmological
perturbation theory.
In this section, we recall the $\delta N$ formula and  its application to 
light fields. Then see how it works for the waterfall field.

\subsection{Initial flat slice}
 
 One may extend  \eq{gij}  to describe a
 generic sequence of fixed-$t$ slices, subject only to the condition that the
initial slice has unit determinant for $g_{ij}$ (called a flat slice) while
the final slice still has uniform density. Since the local scale factor is uniform
on the initial slice, $\zeta$ on the final slice is given by
\be
\zeta(\bfx,t) = \delta N(\bfx,t) \equiv N(\bfx,t) - N(t)
, \dlabel{deltan1} \ee
where $N(\bfx,t)$ is the number of $e$-folds of expansion at a given location,
\be
N(\bfx,t)
\equiv  \int^{t(\rho)}_{t_1} dt' \dot a(\bfx,t')/a(\bfx,t')
, \dlabel{nofxt} \ee
and $N(t)$ is the expansion in the unperturbed universe with scale factor
 $a(t)$.
The expansion is from the initial flat slice at time $t_1$, to the final
uniform-$\rho$ slice at time $t$, and we  suppress the
argument $t_1$ of $N$ because  $\delta N$ is independent of $t_1$.
This formula was given to first order in \cite{ss}
and non-linearly in \cite{lms}.\footnote
{Considering the time-dependence of $\zeta$
during inflation, which was also the focus in \cite{ss}, the first-order formula
is also given in \cite{starob} and its non-linear generalization is given
in \cite{sbb,st}.}

Going to the proper
time $t\sub{pr}$  at each location, \eq{nofxt} becomes
\be
 N(\bfx,t\sub{pr})
\equiv  \int^{ t\sub{pr}(\rho,\bfx)  }_{ t_{1{\rm pr}} (\bfx) } 
 H(\bfx,t\sub{pr}') dt'\sub{pr}
. \dlabel{nofxtsep} \ee
(Without loss of generality one usually takes  $t_{1{\rm pr}}$ to be independent
of $\bfx$.)
For each of the separate universes, $t\sub{pr}$ is the cosmic time appearing the 
Robertson-Walker line element, and $H$ is the Hubble parameter given by
$3\mpl^2 H^2= \rho$.
 Since the final slice has homogeneous  $\rho$, the final value of $H$
is also homogeneous.

\subsection{Initial uniform-density slice}

The contribution to $\zeta$ that is generated between times $t_1$ and $t$ is
\bea
\zeta(\bfx,t)-\zeta(\bfx,t_1) &=& \delta N(\bfx,t,t_1)
\equiv N(\bfx,t,t_1) - N(t,t_1)
\dlabel{deltan2}  \\
 N(\bfx,t,t_1) &\equiv& \int^{t(\rho)}_{t(\rho_1)} dt' \dot a(\bfx,t')/a(\bfx,t')
\dlabel{nofxtt1}
, \eea
where  both the initial and final slices have uniform $\rho$.

Going to proper time, 
\be
N(\bfx,t\sub{pr},t\sub{pr1})  = \int^{ t\sub{pr}(\rho,\bfx) }_{ t\sub{pr}(\rho_1,\bfx) }
H(\bfx,t'\sub{pr})
dt'\sub{pr}
, \dlabel{nofxtt1sep} \ee
with both  the initial and final values of $H$
determined by the relation $3\mpl^2 H^2= \rho$.

\subsection{$\delta N$ for the contribution of light fields}

\subsubsection{Initial flat slice}

Let us now  choose the initial epoch   to be
 a very few Hubble times after  the smoothing scale $L_\zeta$ of $\zeta$
leaves the horizon during inflation, and denote it by
$t_*$. At this stage the light fields are smooth on
the  scale $L_\zeta$, because their classical perturbation exists only
on  scales much bigger than the horizon ($k\ll aH$)
 while their quantum fluctuation has been dropped.

One assumes that   the values of one or more of the light fields,
at the epoch $t_*$,  provide at each location the initial condition for the evolution of
the separate universe. Then, with $N$ defined by \eq{nofxt}, we have in terms of these
values
\be
\zeta(\bfx,t) = \delta N =
N(t,\phi_1^*(\bfx),\phi_2^*(\bfx),\cdots) - \vev{ N(t,\phi_1^*,\phi_2^*,\cdots) }
, \dlabel{deltan11} \ee
where $\phi_i^*$ without an argument denote the unperturbed values, and the perturbations
$\delta\phi_i^*\equiv \phi_i^*(\bfx) - \phi_i^*$ are defined on a flat slice.

In practice,  $\zeta$ is well-approximated by a few terms
of a power-series in the field perturbations:
\be
\zeta(\bfx,t) = \sum N_i(t) \delta\phi_i^*(\bfx) +
+ \frac12 \sum N_{ij}(t) \delta\phi_i^*(\bfx))\delta\phi_j^*(\bfx)
 + \cdots
. \dlabel{deltan22} \ee
In this formula, the partial derivatives $N_i$ etc.\
are  evaluated at the unperturbed point in field space. 

Taking $t$ to be any time after
$\zeta$ has stopped varying, we arrive at the quantity constrained by observation
denoted simply by $\zeta(\bfx)$. Since the observed $\zeta$ is gaussian to high
accuracy,  \eqreff{deltan22} should be dominated on cosmological scales  by the  first term.
This term was given
 in \cite{ss}. The  power series was  given in \cite{lr}, 
 where it was used
to calculate the  bispectrum etc.\ that correspond to non-gaussianity in a couple
models.\footnote
{Considering the time-dependence of $\zeta$
during inflation, which is  also the focus in \cite{ss}, the first-order formula
was  also given in \cite{starob} and its non-linear generalization was  considered
(without a specific application) in \cite{sbb,st}.}

If there is single-field slow-roll inflation and the only contribution to 
$\zeta$ comes from the inflaton $\phi$, the perturbation  $\delta\phi_*$ 
just represents a shift along the inflaton trajectory which does not affect
the subsequent relationship $p(\rho)$.  Then $\zeta$ is time-independent,
and the first term of \eq{deltan22} dominates. To first order in $\delta\phi_*$,
the time shift is $\delta t = \delta\phi_*/\dot\phi_*$, which with $\delta N=-H\delta t$
gives  \eq{zetaofphi}. 
If there is  multi-field slow-roll inflation with two or more fields $\phi_i$,
we can put these fields into
 \eq{deltan22} to arrive at $\zeta(\bfx,t)$ during slow-roll inflation.
For each of the separate universes,
the final epoch should be one of fixed $V\simeq \rho$, which in general will mean
{\em different} final values for the fields.

\subsubsection*{Initial uniform-$\rho$ slice}

So far we have taken $N$ to be defined from an initial flat slice,  leading to
 \eq{deltan22}
with $t_*$ an epoch soon after the smoothing scale $L_\zeta$  leaves the horizon.
This is definitely the formula to use if all of the light field
contributions become significant during inflation (so that we deal with multi-field
inflation). There is another  scenario though, where a different 
approach works better. 
This is the curvaton-type scenario, in which the contribution of some  light
field $\sigma$  starts to become significant only at some epoch $t_1$ after inflation
is over. For this to make sense, the smoothing scale $L_\zeta$ chosen for $\zeta$
has to be outside the horizon at the epoch $t_1$. In principle we should keep
the effect on $\rho$ and $p$ of the modes of $\sigma$ that have entered the horizon
during the era $t_*<t<t_1$, but such modes are likely to have redshifted away (ie.\ 
$|\pa^2 V/\pa \sigma^2|\lsim H^2$ is likely during this era) and they are ignored.
Also, the coupling of $\sigma$ to other fields is supposed to be negligible
during this era. The evolution of $\sigma$ at each location
 is then given by the unperturbed field equation, leading to a relation $\chi_1(\chi_*)$ and
\be
\delta\chi(t_1,\bfx) = \frac{d \chi_1}{d\chi_*} \delta \chi_*
+ \frac12  \frac{d^2 \chi_1}{d\chi_*^2} \( \delta \chi_* \)^2 + \cdots
. \dlabel{zetafriday}  \ee
The field $\sigma(\bfx,t_1)$ is supposed to set the initial condition at $t_1$ so that
\be
\zeta_\chi(\bfx,t) = N_\chi(t,t_1) \delta\chi(t_1,\bfx) + \frac12 N_{\chi\chi}(t,t_1)
\[ \delta\chi(t_1,\bfx) \]^2 + \cdots
. \dlabel{zetachideltan} \ee
 where  a subscript denotes a derivative, $N_\chi\equiv \pa N/\pa \chi$ etc..
\eqs{zetafriday}{zetachideltan}  give $\zeta_\chi$ 
 in terms of the nearly-gaussian perturbation
$\delta\chi_*$ which has the spectrum $(H/2\pi)^2$.

 Following \cite{lr}, this  version of the $\delta N$ formula is generally  used
for the curvaton-type contribution,
 usually with the assumption that $\zeta_\phi$ is negligible so that
$\zeta_\chi$ is the observed quantity. With that assumption,
 there is actually no need to assume slow-roll  inflation;
all one needs is the assumption of  inflation with a specified
Hubble parameter $H(t)$ \cite{curvaton}.

\subsection{$\delta N$ for the waterfall contribution}

\dlabel{viib}

Now we show that the 
 $\delta N$ formula 
\eqreff{nofxtt1sep} provides a direct derivation
of \eq{zetachibest} that we used to calculate the waterfall field contribution 
$\zeta_\chi$.
Our calculation sets $H$ equal to a constant. Also, it 
 begins with a slice of
uniform $\phi$, and  subsequent slices are separated by uniform amounts of proper 
time, so that  $\phi$ remains uniform on these slices and  $\delta\rho = \delta\rho_\chi$.
Using these results, \eq{nofxtt1sep}  indeed gives to first order in $\delta\rho_\chi$:
\bea      
\zeta_\chi(\bfx,t\svev) &=&  H \[ \delta t(\bfx,t\svev) - \delta t(\bfx,t_1) \] \\ 
&\simeq &  - \frac H{\dot\rho(t\svev)}  \delta\rho_\chi(\bfx,t\svev)    
= \frac 13 \frac{\delta\rho_\chi(\bfx,t\svev)}{\rho(t\svev) + p(t\svev)  },
\dlabel{zetachibest2}
 \eea
where $\delta t$ is the time interval from the uniform proper time slicing  to the 
uniform $\rho$ slicing. In the second line we dropped $\delta t(\bfx,t_1)$ which 
as seen in Section \ref{vb}   is negligible.

One might be concerned that setting $H$ equal to a constant is incompatible with
$3\mpl^2 H^2=\rho$. But the non-uniformity of $H$ implied by this relation gives
a negligible contribution to $N$:
\be  \int^{t\svev}_{t_1} \delta H dt = \frac12
\int^{t\svev}_{t_1} H \frac{\delta\rho_\chi}{\rho} dt
<\frac12  H \frac{\delta\rho_\chi(\bfx,t\svev)}{\rho(t\svev)},  \ee  
which  is much less than  \eq{zetachibest2}, justify the assumption that $H$ is    
practically uniform.    
                                          
This use of the $\delta N$ formula looks very different from the one for light fields,
but it can be made to resemble the latter using the local evolution equation  \eq{rough}
for $\chi$. (Recall that this equation is needed if we are to use \eq{zetachibest2} within the 
framework of the present paper, because it is invoked for the calculation of 
 $\rho_\chi$.) With this equation, the time $t\svev(\bfx)$ required to give the 
final energy density a uniform value $\rho(t\svev)$ is given by
\be
\rho(t\svev) = \rho_\phi(t\svev(\bfx)) - \frac{ 3H\mu \sqrt{\tau_1} }2 
\chi^2(\bfx,\tau_1) e^{\frac43 \tau\svev^{3/2}(\bfx) }
. \ee
This gives $N\equiv (H/\mu)\tau\svev(\bfx)$ as a function of $\chi^2(\bfx,\tau_1)$,
and taking its first order perturbation we get
\be \zeta_\chi(\bfx,\rho(t)) = \delta N \simeq 
\left. \frac{d N(\chi^2,\rho) }{ d(\chi^2) }\right|_{\chi^2=\vev{\chi^2}}
 \delta \chi^2(\bfx,t_1) 
.  \dlabel{dNzeta} \ee
Since its input is the same, this will  reproduce \eq{zetachibest2} (hence
\eq{zetachibest})  as   one can check. For instance, in
 the simplest case that $\rho_\phi(\bfx,\tau\svev)$ has a negligible perturbation,
corresponding to $\vev{\dot\chi^2(\tau\svev) }  \gg \dot\phi^2$, it gives
\be
\frac43 \( \mu N/H \) ^{3/2} = \mbox{const} - \ln\[ \chi^2(\bfx,\tau_1) \]
, \ee
so that \eq{dNzeta} gives
\be
\zeta_\chi = \frac H{2\mu \sqrt{\tau\svev} }  \frac{ \delta \chi^2(\bfx,\tau_1) }
{ \vev{\chi^2(\tau_1) } }
= \frac H {2\mu \sqrt{ \tau\svev } } \frac{ \delta \chi^2(\bfx,\tau\svev) }
{ \vev{ \chi^2(\tau\svev) } }
, \ee
which indeed coincides with \eq{zetachibest}
  when $\vev{ \dot\chi^2(\tau\svev) }  \gg \dot\phi^2$.

We emphasize that \eq{dNzeta}, which invokes $\delta\chi^2(\bfx,t_1)$ 
smoothed on a super-horizon
scale, is likely to hold  only during the linear era. There is no reason to think
that the smoothed $\delta \chi^2$ will determine the  evolution of $\zeta$ after that era.
Rather, in order to  calculate that  evolution,
one should calculate $\rho$ and $p$ keeping all modes of $\chi$
and $\phi$ {\em and then} smooth them on a super-horizon scale. 

\section{Earlier calculations}

\dlabel{searlier}

Several papers \cite{fggklt,abc,tachyonic,cpr,ggg,gf,dufaux}
consider the evolution of the waterfall field, without considering 
its effect on the curvature perturbation. Among them, 
\cite{abc,cpr,ggg} invoke the assumptions laid down in Section \ref{ssassumptions}. 
We will not comment on these papers, but turn instead to those that consider the 
curvature perturbation. In doing so, we exclude the 
 non-standard scenario $|m_\chi| \ll H$ considered in  \cite{nonstand} and part of
\cite{bc1,bc2}.

Several papers \cite{bfl,bdd,msw,af} treat the waterfall as two-field inflation
with nonzero $\vev{\chi}\sub{nl}$.
For that to be valid,
we would need to be in the extreme $|m_\chi| \ll H$ scenario 
(considered in parts of \cite{bc1,bc2}) where 
$m_\chi^2(t) \ll H^2$ already when the observable universe leaves the horizon.
The calculation in these papers  is therefore incorrect, and all of them except
\cite{af} reach the incorrect conclusion that $\zeta_\chi$ depends linearly on $\chi$,
with $\calp_{\zeta_\chi}$
nearly scale invariant. (The calculation of \cite{af}, setting without
justification $\vev{\chi}=\sqrt{\vev{\chi^2} } $ reaches a basically correct result
even though the reasoning is incorrect; it makes
$\zeta_\chi$  quadratic in $\delta\chi$, with 
$\calp_{\zeta_\chi}\propto k^3$ and with very very roughly
$\calp_{\zeta_\chi}\sim 1$ at $k=H$.)

Two papers  \cite{fsw,gs} invoke the function $N$ 
 considered at the end of the previous section.
They consider hybrid inflation with the potential $V(\phi)=\frac12m_\phi^2\phi^2$.
The parameters are adjusted to give $f=1$, and are chosen so that the 
waterfall takes a small number of Hubble times, not much less than a Hubble time as we
are assuming. For the present purpose that difference can be ignored, because
the  qualitative evolution of $\chi$ and $\zeta$ is similar in the two cases \cite{mynext}.
With these parameters, $\vev{\dot\chi^2(t\svev})\gg \dot\phi^2$, which as we saw at the end
of Section \ref{vb} means that
the slicing of uniform $\rho$ is also one of uniform $\chi$. The latter  is assumed by
the authors, without proof.
In \cite{fsw} $N$ is calculated numerically,
while in \cite{gs} it is calculated analytically.

Both  papers invoke a quantity, denoted by  $\delta\chi_L$, which is $\chi$
smoothed on the scale of the horizon at the beginning of the waterfall.
In \cite{fsw}, $N$ is regarded as a function of $\chi$ rather than $\chi^2$, and is
 inserted into an expression 
\be
\zeta_\chi(\bfx) =
\left. \frac{ d^2(N)_L }{ d\chi^2 } \right|_{\chi=0}
( \delta \chi_L(\bfx,t_1) )^2 
. \dlabel{fsw} \ee
The subscript attached to $(N)_L$ indicates that a smoothing procedure has been
applied, which is necessary because $N$ itself diverges at $\chi_L=0$.
This divergence occurs because, with their choice of parameters, the energy density
at the epoch $t\svev$ is less than $V_0$. Such an energy density cannot be achieved if
$\chi$ is fixed at 0.

This expression for $\zeta_\chi$ differs from 
\eqreff{dNzeta}
in several ways. First, the expansion is about $\chi=0$ whereas it should be 
about $\chi^2=\vev{\chi^2}$. Second
 $(\delta\chi_L)^2$ (which is smoothed and then squared) should be replaced by  
$\delta \chi^2$ (which is squared and then smoothed).
Third, $(N)_L$ should be replaced
by the unsmoothed quantity $N$. As it happens though, none of these
errors will affect the rough order of magnitude, because with their parameter choice the following
hold; (i) the dominant
mode of $\chi$ is of order  $H$ so that the smoothed quantity
$\delta \chi_L$ will not be very different from $\chi$ itself,
(ii) ignoring the smoothing, 
the Fourier components of $\delta\chi_L^2$ are the same as those of $\delta\chi^2$ 
 (iii)  $N(\chi_L^2)$
is approximately linear and close to $(N(\chi_L^2))_L$ away from the origin.

In  \cite{gs},   the correct \eq{dNzeta} is used, except that $\delta\chi^2$ is replaced by
$\delta\chi_L^2$. Because of this replacement, $P_{\delta\chi^2}(k)\sim P_\chi^2(0)k_*^3$
is replaced by $P_{\chi_L^2}\sim P_\chi^2(0) H^3$ (since $\delta\chi_L$ is smooth on the 
Hubble scale) which means that $\calp_{\zeta_\chi}$ is underestimated by a factor
$(H/k_*)^3$, which with their parameters is roughly $(1/3)^3$.\footnote
{We are denoting the dominant scale of their calculated $\chi$ by $k_*$, the same symbol that we
used for its dominant scale in our calculation.}
  In a footnote of  \cite{gs}, an argument is given which is
intended to justify the dropping of short-scale modes when evaluating the perturbation
in $\chi^2$. They write $\chi=\chi_L + \chi_S$, and to define
 $\chi_S$  the universe is divided into boxes with size 
$\sim H\mone$. Within each box $\chi_S$ is defined as the usual 
 Fourier Series valid within that box, and all Fourier coefficients are generated as
usual from the vacuum fluctuation. This proposal is not viable, 
because 
for a typical realization of the ensemble (presumed to correspond to the 
observed universe) $\chi$  defined in this way 
has a large discontinuity at the boundary of each box
in violation of the field equation.

In the Appendix of \cite{gs}, $\zeta_\chi$ is calculated by integrating
$\dot\zeta$, but they get a result that
is bigger than the one from $\delta N$, by a factor 
$\ln[\vev{ \dot\chi^2(t\svev) }/\dot\phi^2(t\svev) ]$. The discrepancy arises 
 because they identify $\delta p\sub{nad}$ with  with $\delta p_\chi$, which
makes $\dot \zeta$ constant in the regime $\vev{\dot\chi^2(t)}
\gg \dot\phi^2$. As  we see from \eq{pnad2},  $\delta p\sub{nad}$ in that   regime
actually decays exponentially.

The papers we considered so far contain errors of principle. We turn now to
papers \cite{ev,bc2}\footnote
{The  paper \cite{bc2} builds on \cite{bc1} and corrects some errors in the latter.} 
 where the method is in principle
correct,  but leads to excessively complicated equations. Presumably because of this
complication, the expressions for $\zeta_\chi$ in \cite{ev,bc2} are  not correct.
The method is to consider equations that include, 
in addition to the perturbations  included in our  equations,
 the second order metric perturbation as well as the 
the product $\chi\delta \phi$.  We demonstrated in Section \ref{sduf} that the latter
has a negligible effect and we pointed out in Section \ref{sszeta} that the former will
also have a negligible effect. The motivation for including these small extra terms
seems to be the following. First, 
 field perturbations are defined as perturbations away from
zero, so that $\chi$ is regarded as a perturbation. Second, 
 `first order' equations are by definition linear in the
field perturbations, so that in this order
 $\chi$ gives no contribution to 
$\rho$ or $p$.  Finally, the next (`second') order is defined to 
involve equations containing terms of the form $x^2$, $xy$ and $y^2$, where $x$
is a first-order field perturbation and $y$ is a first-order metric perturbation.
It is the inclusion of the  $y^2$ terms that causes the main complication in the
equations.

As has been noticed elsewhere \cite{lr,v,bc2}, the  
 expression for $\zeta_\chi$ in \cite{ev} is manifestly incorrect because it 
is  not a local function of $\chi$.\footnote
{Since the equations in \cite{ev} are erroneous, we will not consider \cite{ejmmv}
which uses them to consider  a metric perturbation related to $\zeta$.}
 The expression in \cite{bc2}
(Eq.~(57)) is local, but still not correct. In particular, it should in the 
 regime $6Ht\svev \ll 1$ 
 coincide (to first order in $\delta p_\chi$) with
\eq{zetachibest}, whereas it differs by having $-m^2$ instead of our $-\mu^2\tau
\equiv m^2(\phi)$  (and by a factor 2).
Further calculations using the same
expression were done in \cite{hc}. 

\section{Conclusion}
\dlabel{sconc}

In this  paper we have considered 
$\zeta_\chi$, the amount by which $\zeta$ changes during the 
 linear evolution of the waterfall field. 
We have calculated its spectrum  in  the regime of parameter space 
in which slow-roll inflation continues during the linear era, with 
 the waterfall mass-squared decreasing  linearly and the waterfall takes much less 
$Ht\svev \ll 1$ Hubble times. The spectrum, given by \eq{calpzchi}, 
is proportional to $k^3$ and amply  satisfies the black hole bound $\calpz\lsim 10\mtwo$
on  those super-horizon scales $k\ll H$ for which $\zeta$ is defined.
The full parameter space for standard hybrid inflation will be explored elsewhere
\cite{mynext}, 
using instead of $\zeta$ the quantity $\calr$ mentioned before \eq{beforefdef}.

We made 
 the  fundamental assumption, shared by all previous authors except those
of \cite{dufaux},  that the waterfall does indeed begin  with a the  linear era. 
 We have justified that assumption using the
work of  \cite{dufaux}, who invoke in turn the calculation of \cite{tachyonic}. 
But  the justification is not quite complete, because the
latter  calculation invokes a nonzero value of $\vev{\chi^2}$
at the start of the waterfall, coming from the vacuum fluctuation, whereas 
 we have set that value to zero.

This brings us to  an   issue of principle, 
 namely the regularization  of uv divergences. We adopted the simplest procedure,
of dropping at a given epoch all  Fourier
modes (of the waterfall and inflaton fields) that are in the quantum as opposed
to the classical regime. 
That  procedure will violate at some level the  
 energy continuity equation, which follows from the complete field equations
including all modes. With the assumptions laid down in Section 
\ref{ssassumptions}, the violation will be negligible  because
the dominant classical modes (with 
 $k\sim k_*$) appear as soon as $\tau$ becomes significantly bigger
than 1; as a result, 
the subsequent creation of classical modes, 
which violates the energy continuity equation, can be 
ignored. But  there seems to be no guarantee that the violation 
 will be negligible in all
cases where a classical field is created from the vacuum.

More seriously, the procedure of dropping the quantum modes
will not make sense at all for   the 
region of parameter space in which the
 waterfall field fails to enter the classical regime.  (We identified one such a region
in Section \ref{sgut} working with our particular assumptions.)
 In that region,
the waterfall becomes an essentially quantum phenomenon requiring a 
formalism quite different from the one that we presented.

The procedure of dropping the quantum regime is crude, but  something certainly has to be
done because the energy density, pressure and mean-square fields diverge if the entire
quantum regime is kept. Indeed, an ultra-violet  cutoff at  some scale  $k/a=\Lambda$
 much bigger than the particle masses
gives a  constant  energy density and pressure $\rho=3 p=\Lambda^4/16\pi^2$ which violates
 the energy 
continuity equation \cite{magg}.\footnote 
{Contrary to an  assumption that has been widely adopted following the
seminal review \cite{weinberg}, this is not a contribution to the cosmological
constant which would require $p=-\rho$, satisfying the energy continuity equation.} 
Several  alternatives to simply 
dropping the quantum regime have been proposed that can apply to hybrid
inflation (for example \cite{abc,cpr,ggg,regul} in addition to the procedure
of \cite{tachyonic} that we mentioned before)
 and  it is not clear how to choose between them if they give different results.

\section{Acknowledgments}
The author acknowledges useful correspondence with the most of the authors of 
\cite{dufaux,bfl,bdd,msw,af,bc1,fsw,gs}, and 
support from
 EU grants MRTN-CT-2006-035863 and UNILHC23792, 
European Research and Training Network (RTN) grant. 



\begin{thebibliography}{999}


\bibitem{bhbound}
 B.~J.~Carr, K.~Kohri, Y.~Sendouda and J.~Yokoyama,
``New cosmological constraints on primordial black holes,''
  Phys.\ Rev.\  D {\bf 81} (2010) 104019.


\bibitem{efg}
  R.~Easther, R.~Flauger and J.~B.~Gilmore,
``Delayed Reheating and the Breakdown of Coherent Oscillations,''
  JCAP {\bf 1104} (2011) 027, and earlier work cited there.

\bibitem{ouranis}
 K.~Dimopoulos,
  ``Can a vector field be responsible for the curvature perturbation in the
  universe?,''
  Phys.\ Rev.\  D {\bf 74} (2006) 083502;
S.~Yokoyama and J.~Soda,
  ``Primordial statistical anisotropy generated at the end of inflation,''
  JCAP {\bf 0808} (2008) 005;
 K.~Dimopoulos, M.~Karciauskas, D.~H.~Lyth and Y.~Rodriguez,
  ``Statistical anisotropy of the curvature perturbation from vector field
  perturbations,''
  JCAP {\bf 0905} (2009) 013;
 M.~a.~Watanabe, S.~Kanno and J.~Soda,
  ``The Nature of Primordial Fluctuations from Anisotropic Inflation,''
  Prog.\ Theor.\ Phys.\  {\bf 123} (2010) 1041.

\bibitem{ourbh}
K.~Kohri, D.~H.~Lyth and A.~Melchiorri,
  ``Black hole formation and slow-roll inflation,''
  JCAP {\bf 0804}, 038 (2008).

\bibitem{standardph}
 J.~H.~Traschen and R.~H.~Brandenberger,
  ``Particle production during out-of-equilibrium phase transitions,''
  Phys.\ Rev.\  D {\bf 42} (1990) 2491;
  L.~Kofman, A.~D.~Linde and A.~A.~Starobinsky,
  ``Reheating after inflation,''                                               
  Phys.\ Rev.\ Lett.\  {\bf 73} (1994) 3195.

\bibitem{ourfirstph}
 A.~R.~Liddle, D.~H.~Lyth, K.~A.~Malik and D.~Wands,
  ``Super-horizon perturbations and preheating,''
  Phys.\ Rev.\  D {\bf 61}, 103509 (2000).

\bibitem{kt}
 M.~Kawasaki, T.~Takayama, M.~Yamaguchi and J.~Yokoyama,
  ``Power Spectrum of the Density Perturbations From Smooth Hybrid New
  Inflation Model,''
  Phys.\ Rev.\  D {\bf 74} (2006) 043525.

\bibitem{smoothhybrid}
  E.~D.~Stewart,
  ``Mutated hybrid inflation,''
  Phys.\ Lett.\ B {\bf 345}, 414 (1995);
 G.~Lazarides and C.~Panagiotakopoulos,
  ``Smooth hybrid inflation,''
  Phys.\ Rev.\ D {\bf 52} (1995) 559.

\bibitem{book}
 D. H. Lyth and A. R. Liddle, {\it The primordial density perturbation},
Cambridge University Press, 2009; 
{\tt http://astronomy.sussex.ac.uk/~andrewl/PDP/errata.pdf};
{\tt http://astronomy.sussex.ac.uk/~andrewl/PDP/extensions.pdf}.

\bibitem{myaxion}
  D.~H.~Lyth,
  ``Axions And Inflation: Sitting In The Vacuum,''
  Phys.\ Rev.\  D {\bf 45} (1992) 3394.

\bibitem{curvaton}
  D.~H.~Lyth and D.~Wands,
``Generating the curvature perturbation without an inflaton,''
  Phys.\ Lett.\  B {\bf 524} (2002) 5;

\bibitem{curvaton2}
 T.~Moroi and T.~Takahashi,
``Effects of cosmological moduli fields on cosmic microwave background,''
  Phys.\ Lett.\  B {\bf 522} (2001) 215
  [Erratum-ibid.\  B {\bf 539} (2002) 303].

\bibitem{klv}
  K.~Kohri, D.~H.~Lyth and C.~A.~Valenzuela-Toledo,
``On the generation of a non-gaussian curvature perturbation during
preheating,''
  JCAP {\bf 1002} (2010) 023.

\bibitem{gwbm}
  C.~Gordon, D.~Wands, B.~A.~Bassett, R.~Maartens,
``Adiabatic and entropy perturbations from inflation,''
  Phys.\ Rev.\  {\bf D63}, 023506 (2001).

\bibitem{andreihybrid}
 A.~D.~Linde,
  ``Axions in inflationary cosmology,''                                       
  Phys.\ Lett.\ B {\bf 259} (1991) 38.

\bibitem{myfirstaxion}
 D.~H.~Lyth,
  ``A limit on the inflationary energy density from axion isocurvature fluctuations,''
  Phys.\ Lett.\  B {\bf 236} (1990) 408.

\bibitem{axionstring}
  D.~H.~Lyth and E.~D.~Stewart,
  ``Axions And Inflation: String Formation During Inflation,''
  Phys.\ Rev.\  D {\bf 46} (1992) 532.

\bibitem{earlyhybrid}
  B.~A.~Ovrut and P.~J.~Steinhardt,
``Inflationary Cosmology And The Mass Hierarchy In Locally Supersymmetric
Theories,''
  Phys.\ Rev.\ Lett.\  {\bf 53} (1984) 732;
  K.~Enqvist and D.~V.~Nanopoulos,
``Primordial Two Component Inflation,''
  Nucl.\ Phys.\  B {\bf 252} (1985) 508.

\bibitem{ourhybrid}
E.~J.~Copeland, A.~R.~Liddle, D.~H.~Lyth, E.~D.~Stewart and D.~Wands,
  ``False vacuum inflation with Einstein gravity,''
  Phys.\ Rev.\  D {\bf 49}, 6410 (1994).

\bibitem{gutinf}
 G.~R.~Dvali, Q.~Shafi and R.~K.~Schaefer,
  ``Large scale structure and supersymmetric inflation without fine tuning,''
  Phys.\ Rev.\ Lett.\  {\bf 73}, 1886 (1994).

\bibitem{supernatural}
 L.~Randall, M.~Soljacic and A.~H.~Guth,
  ``Supernatural inflation: inflation from Supersymmetry with No (Very) Small  
  Parameters,''                                                                
  Nucl.\ Phys.\  B {\bf 472}, 377 (1996).

\bibitem{runningmass}
 E.~D.~Stewart,
``Flattening the inflaton's potential with quantum corrections,''
Phys.\ Lett.\ B {\bf 391} (1997) 34;
E.~D.~Stewart,
``Flattening the inflaton's potential with quantum corrections. II,''
Phys.\ Rev.\ D {\bf 56} (1997) 2019.

\bibitem{dbi}
 M.~Alishahiha, E.~Silverstein and D.~Tong,
  ``DBI in the sky,''
  Phys.\ Rev.\  D {\bf 70} (2004) 123505.

\bibitem{lbound}
 D.~H.~Lyth,
  ``What would we learn by detecting a gravitational wave signal in the  cosmic
  microwave background anisotropy?,''
  Phys.\ Rev.\ Lett.\  {\bf 78}, 1861 (1997);
 L.~Boubekeur and D.~H.~Lyth,
  ``Hilltop inflation,''
  JCAP {\bf 0507}, 010 (2005).

\bibitem{al}
 L.~Alabidi and D.~H.~Lyth,
  ``Inflation models and observation,''
  JCAP {\bf 0605} (2006) 016.

\bibitem{inverted}
  D.~H.~Lyth and E.~D.~Stewart,
  ``More varieties of hybrid inflation,''
  Phys.\ Rev.\ D {\bf 54}, 7186 (1996).

\bibitem{firstorder}
 A.~D.~Linde,
``Eternal extended inflation and graceful exit from old inflation without
Jordan-Brans-Dicke,''
  Phys.\ Lett.\  B {\bf 249} (1990) 18;
 F.~C.~Adams and K.~Freese,
``Double field inflation,''
  Phys.\ Rev.\  D {\bf 43} (1991) 353.

\bibitem{areview}
  A.~R.~Liddle and D.~H.~Lyth,
  ``The Cold dark matter density perturbation,''
  Phys.\ Rept.\  {\bf 231}, 1 (1993).

\bibitem{bhend}
 D.~H.~Lyth, K.~A.~Malik, M.~Sasaki and I.~Zaballa,
``Forming sub-horizon black holes at the end of inflation,''
  JCAP {\bf 0601} (2006) 011;
 I.~Zaballa, A.~M.~Green, K.~A.~Malik and M.~Sasaki,
``Constraints on the primordial curvature perturbation from primordial black
holes,''
  JCAP {\bf 0703} (2007) 010.

\bibitem{mybox}
 D.~H.~Lyth,
  ``The curvature perturbation in a box,''
  JCAP {\bf 0712} (2007) 016.

\bibitem{dufaux}
 J.~F.~Dufaux, G.~N.~Felder, L.~Kofman and O.~Navros,
  ``Gravity Waves from Tachyonic Preheating after Hybrid Inflation,''
  JCAP {\bf 0903} (2009) 001.

\bibitem{tachyonic}
 G.~N.~Felder, L.~Kofman and A.~D.~Linde,
  ``Tachyonic instability and dynamics of spontaneous symmetry breaking,''
  Phys.\ Rev.\  D {\bf 64}, 123517 (2001).

\bibitem{mynext}
D.~H.~Lyth, in preparation.


\bibitem{ss}
  M.~Sasaki and E.~D.~Stewart,
``A General Analytic Formula For The Spectral Index Of The Density
Perturbations Produced During Inflation,''
  Prog.\ Theor.\ Phys.\  {\bf 95} (1996) 71.

\bibitem{lms}
  D.~H.~Lyth, K.~A.~Malik and M.~Sasaki,
``A general proof of the conservation of the curvature perturbation,''
  JCAP {\bf 0505} (2005) 004.

\bibitem{starob}
 A.~A.~Starobinsky, ``Multicomponent de Sitter                           
(inflationary) stages and the generation of perturbations``, 
Pis'ma Zh. Eksp. Teor. Fiz. {\bf      
42}, 124 (1985) [JETP Lett. {\bf 42}, 152 (1985)].

\bibitem{sbb}
  D.~S.~Salopek, J.~R.~Bond and J.~M.~Bardeen,
``Designing Density Fluctuation Spectra in Inflation,''
  Phys.\ Rev.\  D {\bf 40} (1989) 1753.

\bibitem{st}
  M.~Sasaki and T.~Tanaka,
  ``Super-horizon scale dynamics of multi-scalar inflation,''
  Prog.\ Theor.\ Phys.\  {\bf 99} (1998) 763.

\bibitem{lr}
  D.~H.~Lyth and Y.~Rodriguez,
``The inflationary prediction for primordial non-gaussianity,''
  Phys.\ Rev.\ Lett.\  {\bf 95} (2005) 121302.

\bibitem{fggklt}
  G.~N.~Felder, J.~Garcia-Bellido, P.~B.~Greene, L.~Kofman, A.~D.~Linde and I.~Tkachev,
``Dynamics of symmetry breaking and tachyonic preheating,''
  Phys.\ Rev.\ Lett.\  {\bf 87} (2001) 011601.

\bibitem{abc}
 T.~Asaka, W.~Buchmuller and L.~Covi,
``False vacuum decay after inflation,''
  Phys.\ Lett.\  B {\bf 510} (2001) 271.

\bibitem{cpr}
 E.~J.~Copeland, S.~Pascoli and A.~Rajantie,
  ``Dynamics of tachyonic preheating after hybrid inflation,''
  Phys.\ Rev.\  D {\bf 65}, 103517 (2002).

\bibitem{ggg}
J.~Garcia-Bellido, M.~Garcia Perez and A.~Gonzalez-Arroyo,
  ``Symmetry breaking and false vacuum decay after hybrid inflation,''
  Phys.\ Rev.\  D {\bf 67} (2003) 103501;

\bibitem{gf}
 J.~Garcia-Bellido and D.~G.~Figueroa,
``A stochastic background of gravitational waves from hybrid preheating,''
  Phys.\ Rev.\ Lett.\  {\bf 98} (2007) 061302;

\bibitem{nonstand}
 L.~Randall, M.~Soljacic and A.~H.~Guth,
``Supernatural inflation: Inflation from supersymmetry with no (very) small
parameters,''
  Nucl.\ Phys.\  B {\bf 472} (1996) 377;
 J.~Garcia-Bellido, A.~D.~Linde and D.~Wands,
  Phys.\ Rev.\  D {\bf 54} (1996) 6040;
 D.~Parkinson, S.~Tsujikawa, B.~A.~Bassett and L.~Amendola,
``Testing for double inflation with WMAP,''
  Phys.\ Rev.\  D {\bf 71} (2005) 063524;
  S.~Tsujikawa, D.~Parkinson and B.~A.~Bassett,
``Correlation - consistency cartography of the double inflation
landscape,''
  Phys.\ Rev.\  D {\bf 67} (2003) 083516;
A.~A.~Abolhasani, H.~Firouzjahi and M.~H.~Namjoo,
  Class.\ Quant.\ Grav.\  {\bf 28} (2011) 075009.

\bibitem{bc1}
 N.~Barnaby and J.~M.~Cline,
``Nongaussian and nonscale-invariant perturbations from tachyonic preheating
in hybrid inflation,''
  Phys.\ Rev.\  D {\bf 73} (2006) 106012.

\bibitem{bc2}
  N.~Barnaby and J.~M.~Cline,
  ``Nongaussianity from Tachyonic Preheating in Hybrid Inflation,''
  Phys.\ Rev.\  D {\bf 75} (2007) 086004. 

\bibitem{bfl}
 R.~H.~Brandenberger, A.~R.~Frey and L.~C.~Lorenz,
  ``Entropy Fluctuations in Brane Inflation Models,''
  Int.\ J.\ Mod.\ Phys.\  A {\bf 24} (2009) 4327.

\bibitem{bdd}
  R.~H.~Brandenberger, K.~Dasgupta and A.~C.~Davis,
  ``A Study of Structure Formation and Reheating in the D3/D7 Brane Inflation
  Model,''
  Phys.\ Rev.\  D {\bf 78} (2008) 083502.

\bibitem{msw}
 D.~Mulryne, D.~Seery and D.~Wesley,
  ``Non-Gaussianity constrains hybrid inflation,''
  arXiv:0911.3550 [astro-ph.CO].

\bibitem{af}
  A.~A.~Abolhasani and H.~Firouzjahi,
``No Large Scale Curvature Perturbations during Waterfall of Hybrid
Inflation,''
  Phys.\ Rev.\  D {\bf 83} (2011) 063513.

\bibitem{fsw}  J.~Fonseca, M.~Sasaki and D.~Wands,
  ``Large-scale Perturbations from the Waterfall Field in Hybrid Inflation,''
  JCAP {\bf 1009} (2010) 012.

\bibitem{gs}
 J.~O.~Gong and M.~Sasaki,
``Waterfall field in hybrid inflation and curvature perturbation,''
  JCAP {\bf 1103} (2011) 028.

\bibitem{ev}
 K.~Enqvist and A.~Vaihkonen,
  ``Non-Gaussian perturbations in hybrid inflation,''
  JCAP {\bf 0409} (2004) 006.

\bibitem{v}
  A.~Vaihkonen,
  ``Comment on non-Gaussianity in hybrid inflation,''
  arXiv:astro-ph/0506304.

\bibitem{ejmmv}
 K.~Enqvist, A.~Jokinen, A.~Mazumdar, T.~Multamaki and A.~Vaihkonen,
  ``Non-gaussianity from instant and tachyonic preheating,''
  JCAP {\bf 0503} (2005) 010;
 K.~Enqvist, A.~Jokinen, A.~Mazumdar, T.~Multamaki and A.~Vaihkonen,
  ``Cosmological constraints on string scale and coupling arising from
  tachyonic instability,''
  JHEP {\bf 0508} (2005) 084.

\bibitem{hc}
J.~M.~Cline and L.~Hoi,
  ``Inflationary potential reconstruction for a WMAP running power  spectrum,''
  JCAP {\bf 0606} (2006) 007.

\bibitem{magg}
 M.~Maggiore,
``Zero-point quantum fluctuations and dark energy,''
  Phys.\ Rev.\  D {\bf 83} (2011) 063514;
 G.~Mangano,
  ``Shadows of trans-planckian physics on cosmology and the role of the
  zero-point energy density,''
  Phys.\ Rev.\  D {\bf 82} (2010) 043519;

\bibitem{weinberg}
 S.~Weinberg,
  ``The cosmological constant problem,''        
  Rev.\ Mod.\ Phys.\  {\bf 61} (1989) 1.

\bibitem{regul}
 J.~Baacke, L.~Covi and N.~Kevlishvili,
  ``Coupled scalar fields in a flat FRW universe: renormalization,''
  JCAP {\bf 1008} (2010) 026;
 S.~Borsanyi and U.~Reinosa,
  ``Renormalized nonequilibrium quantum field theory: scalar fields,''
  Phys.\ Rev.\  D {\bf 80} (2009) 125029;
 D.~Cormier, K.~Heitmann and A.~Mazumdar,
 ``Dynamics of coupled bosonic systems with applications to preheating,''
  Phys.\ Rev.\  D {\bf 65} (2002) 083521;
  D.~Boyanovsky, D.~Cormier, H.~J.~de Vega, R.~Holman, 
A.~Singh and M.~Srednicki,
  ``Scalar field dynamics in Friedman Robertson Walker spacetimes,''
  Phys.\ Rev.\  D {\bf 56} (1997) 1939.
 S.~Weinberg,
``Ultraviolet Divergences in Cosmological Correlations,''
  Phys.\ Rev.\  D {\bf 83} (2011) 063508.

\end{thebibliography}
\end{document}